\documentclass[IEEEtrans]{IEEEtran}
\ifx\pdfoutput\undefined
\usepackage{graphicx}
\else
\usepackage[pdftex]{graphicx}
\fi
\usepackage{empheq}
\usepackage{url}
\usepackage{array}
\usepackage{stfloats}
\usepackage{amssymb}
\usepackage{amsmath}
\usepackage{amsthm}
\usepackage[noadjust]{cite}
\usepackage[colorlinks]{hyperref}
\usepackage{enumerate}
\usepackage{epstopdf}
\usepackage{color}
\usepackage{multirow}

\DeclareGraphicsExtensions{.pdf,.png,.jpg}

\DeclareMathOperator{\tr}{tr}
\newcommand\captionof[1]{\def\@captype{#1}\caption}
\begin{document}
\title{Symbol-level and Multicast Precoding for Multiuser Multiantenna Downlink: A Survey, Classification and Challenges}
 \author{
  \IEEEauthorblockN{ Maha~Alodeh,~\IEEEmembership{Member, IEEE}, Danilo~Spano ,~\IEEEmembership{Student Member, IEEE}, Ashkan~Kalantari ~\IEEEmembership{Member, IEEE}, Christos~Tsinos ~\IEEEmembership{Member, IEEE}, Dimitrios~Christopoulos ~\IEEEmembership{Member, IEEE}, Symeon~Chatzinotas \IEEEmembership{Senior~Member,~IEEE,}
 Bj\"{o}rn Ottersten, \IEEEmembership{Fellow Member,~IEEE}}\\
\thanks{Maha Alodeh, Danilo Spano, Ashkan Kalantari, Christos Tsinos, Symeon Chatzinotas and  Bj\"{o}rn Ottersten are with Interdisciplinary Centre for Security Reliability and Trust (SnT) at the University
	of Luxembourg, Luxembourg, e-mails:\{ maha.alodeh, danilo.spano, ashkan.kalantari, christos.tsinos, symeon.chatzinotas, and bjorn.ottersten@uni.lu\}, Dimitrios Christopoulos is with Newtec Satcom, Belgium, e-mail:\{dchr@newtec.eu\}. \newline
	This work is supported by Fond National de la Recherche Luxembourg (FNR) projects: SATellite SEnsor NeTworks for Spectrum Monitoring (SATSENT), Spectrum Management and Interference Mitigation in Cognitive Hybrid Satellite Network (SEMIGOD),  Spectrum Management and Interference Mitigation in Cognitive Hybrid Satellite Network (INWIPNET), Energy and Complexity Efficient Millimeter-wave Large-Array Communications (ECLECTIC), Broadband/Broadcast Convergence through Intelligent Caching in 5G Satellite Networks (no. 10079323),  and H2020 Project Shared Access terrestrial-satellite backhaul Network enabled by Smart Antennas (SANSA). 
	}  
}
 
 \maketitle

\begin{abstract}
Precoding has been conventionally considered as an effective means of mitigating the interference and efficiently exploiting the available in the multiantenna downlink channel, where multiple users are simultaneously served with independent information over the same channel resources. The early works in this area were focused on transmitting an individual information stream to each user by constructing weighted linear combinations of symbol blocks (codewords). However, more recent works have moved beyond this traditional view by: i) transmitting distinct data streams to groups of users and ii) applying precoding on a symbol-per-symbol basis. In this context, the current survey presents a unified view and classification of precoding techniques with respect to two main axes: i) the switching rate of the precoding weights, leading to the classes of block- and symbol-level precoding, ii) the number of users that each stream is addressed to, hence unicast-/multicast-/broadcast- precoding. Furthermore, the classified techniques are compared through representative numerical  results to demonstrate their relative performance and uncover fundamental insights. Finally, a list of open theoretical problems and practical challenges are presented to inspire further research in this area.\footnote{The concepts of precoding and beamforming are used interchangeably throughout the paper.}
\end{abstract}
\section{Introduction}\label{introduction}
\IEEEPARstart{P}{recoding} has been a very prolific research area in recent years due to the promise of breaking the throughput gridlock of many interference-limited systems. 
The precoding performance gains originate in the combination of aggressive frequency reuse and suitable interference management techniques. Early works have focused in single-cell scenarios where the main limitation is intra-cell interference \cite{roy,Bjornson_SPM,Gesbert_SPM_MUMIMO,Tutorial_kountouris,clerckx_wimax,Per_Zetterberg}, while later works have also considered multi-cell and heterogeneous networks where inter-system interference \cite{Gesbert_JSAC_cellular,Bassoy_2017,Schwarz_2014,Li_2014} had to be considered as well. It should be noted that precoding has found applications in many practical communication systems, such as terrestrial cellular \cite{Gesbert_JSAC_cellular,Bassoy_2017,Schwarz_2014,Li_2014,Book_comp}, satellite \cite{Arapoglu_MIMO_Satellite,SnT_MIMO_Satellite,Arapoglou_2015}, Digital Subscriber Line (DSL) \cite{DSL_MIMO}, powerline\cite{Power_line_Communications,Berger_2015}, and visible light communications \cite{Wang_2015,Shen_2016_vlc,Cai_2016_vlc}. However, in order to provide a unifying view, this paper does not consider the peculiarities of each application area (e.g. channel, network architecture) but it rather focuses on a general communication model which can encompass the majority of precoding techniques. 

Focusing on interference, this 
is one of the crucial and limiting factors in wireless networks. The
concept of exploiting the users' spatial separation has been a fertile research domain for more than two
decades \cite{roy,Per_Zetterberg}. This can be implemented by adding multiple antennas at one or both
communication sides. Multiantenna transceivers empower communication systems with more degrees
of freedom that can boost the performance if the multiuser interference is mitigated
properly. In this context, the term precoding can be broadly defined as the design  of the transmitted signal to \textit{efficiently} deliver \textit{the desired data stream at each user} exploiting the multiantenna spatial degrees  of freedom, data and channel state information while \textit{limiting the inter-stream interference}.

In this survey, we use two major axes of classification depending on: 
\begin{itemize}
\item The \textbf{switching rate}: how often the precoding coefficients are updated,
\item The \textbf{group size}: the number of targeted users per information stream.
\end{itemize}
In the first classification, we differentiate between \textbf{block-level} and \textbf{symbol-level} precoding. In the former, the precoding coefficients are applied across block of symbols (or codewords), whereas in the latter they are applied on a symbol basis, i.e. switching with the baud rate. The second classification axis differentiates 
according to the requested
service, namely among \textbf{broadcast, unicast, and multicast}. The first service type is known as broadcast, in which a transmitter has a common message to be sent to multiple receivers. In physical layer research,  this service has been studied under the term of physical layer multicasting (i.e. \textit{PHY multicasting}) \cite{Sidiropoulos:2006}-\cite{multicast-jindal}. Since a single data stream
is sent to all receivers, there is no multiuser interference. However, precoding can still be used to improve the quality of service ($\mathrm{QoS}$) across all users. 
The second service type is known as unicast, in which a transmitter
has an individual message for each receiver. Due to the nature of the wireless medium
and the use of multiple antennas, multiple simultaneous unicast transmissions are possible. In these cases, multiple streams are
simultaneously sent, which motivates precoding techniques that mitigate the
multiuser interference. From an information theoretic point of view, this
service type has been studied using the broadcast channel \cite{caire_shitz_TIT}. 
Finally, the multicast service refers to the case where multiple messages are transmitted simultaneously
but each message is addressed to a group of users. This case is also known as multigroup
multicast precoding \cite{Karipidis2005CAMSAP,Karipidis2007,Karipidis2008,Christopoulos2014TWCOM,Christopoulos2014_ICC,Christopoulos2014_TSP,Christopoulos2015_SPAWC}
\footnote{It should be noted that alternative transmission strategies, such as rate-splitting and channels with both individual and common data will not be covered in this survey.}. The classification methodology is further detailed in Section \ref{Classification}.

\subsubsection{Outline and Notation}
This paper starts with introducing the scope of this survey by describing the communications model and the classification methodology in Section \ref{introduction}.  Then, it proceeds to the preliminaries in Section \ref{Preliminaries}. Section \ref{sec:block:level:pre} describes in detail the fundamentals of block-level multicast precoding. Section \ref{sec:slp:unicast} states the connection between the directional modulation and symbol-level precoding. Comparative studies between symbol-level and block level precoding as well as between block-level unicast, broadcast and multicast are conducted in Section \ref{Comparative_Study}.  Some challenges and open problems are thoroughly discussed \ref{Open Problems}. Finally, Section \ref{conclusion} concludes the survey.

\textbf{Notation}:  We use boldface upper and lower case letters for
matrices and column vectors, respectively. $(\cdot)^H$, $(\cdot)^*$ and $(\cdot)^\dagger$
stand for the Hermitian transpose, conjugate and transpose of $(\cdot)$ respectively. $\mathbb{E}(\cdot)$ and $\|\cdot\|$ denote the statistical expectation and the Euclidean norm. $\angle(\cdot)$, $|\cdot|$ are the angle and magnitude  of $(\cdot)$ respectively. $\mathcal{R}(\cdot)$, $\mathcal{I}(\cdot)$
are the real and the imaginary part of $(\cdot)$. Finally, $\tr(\cdot)$ denotes the trace $(\cdot)$ and $[\cdot]_{m,n}$ denotes the element in the row $m$ and column $n$ of $[\cdot]$.
\subsection{Communication Model}
Let us assume that a base station (BS) equipped with $N$ transmit antennas and wishes to transmit $M$ number of symbol streams to $K$ single-antenna users. Adopting a baseband discrete memoryless model, the received signal at the $k$th user for the symbol slot $t$ can be written as:
\begin{eqnarray}
{y_k}[t]=\mathbf{h}^\dagger_k\mathbf{x}[t]+{z}_k[t],
\end{eqnarray}
where $\mathbf{h}_k$ is an $N\times1$ complex vector representing the channel of the $k$th user, $\mathbf{x}[t]$ is an $N\times1$ complex vector representing the output signal from the $N$ transmit antennas and ${z}_k[t]$ is a complex scalar representing the Additive White Gaussian Noise (AWGN).  

The above communication model can be equivalently written in a vector form as:
\begin{equation}
\mathbf{y}[t]=\mathbf{H}^\dagger\mathbf{x}[t]+\mathbf{z}[t],
\end{equation}
where $\mathbf{y}$ is a $K\times1$ complex vector representing the received signal at all $K$ users, $\mathbf{H}=[\mathbf{h}_1 \ldots \mathbf{h}_K]$ is an $N\times K$ complex matrix representing the system channel matrix and $\mathbf{z}$ is a $K\times 1$ complex vector representing the AWGN for all $K$ users.

It should be noted that in the context of this paper, we assume that each symbol stream is divided into blocks of $T$ symbols, while the channel matrix $\mathbf{H}$ remains constant for each block of symbols. In this context, $\mathbf{S}=[\mathbf{s}_1 \ldots \mathbf{s}_K]^\dagger$ is an $M \times T$  complex matrix aggregating the $T \times 1$ input symbol vectors $\mathbf{s}_k$ for each user or group $k$, which are assumed uncorrelated in time and space and having unit average power $\mathbb{E}_t[\mathbf{s}^H_k\mathbf{s}_k]=1$. Analogously, the $N \times T $ matrix $\mathbf{X}$ represents the block of output signals. In terms of system dimensions, we assume that $N \geq K$ and $K \geq M$. In case $K > M$, we assume that the users can be split in $M$ equal groups of $G=K/M$ users per group.
 
\begin{table}
	\begin{center}
		\begin{tabular}{|p{1.3cm}|p{7cm}|}
			\hline
			Parameter& Definition\\
			\hline
			N& Number of transmit antennas \\
			\hline
			K& Number of single antenna users\\
			\hline
			G& Number of groups\\
			\hline
			M& Number of symbol streams, for unicast $M=K$, multicast $M=G$ and broadcast $M=1$\\
			\hline
			T& The number of transmitted symbols in each block\\
			\hline
			$\mathbf{h}_k$& The channel between the BS and user $k$\\
			\hline
			$\mathbf{s}_k[t]$& The data stream (i.e. the set of symbols) dedicated to $k$ user or group/user\\
            \hline
         $\mathbf{S}$& complex
matrix aggregating the data streams to be sent to all users in the coherence time \\
            \hline
            $\mathbf{x}[t]$& The output vector from the antennas\\
            \hline
            $\mathbf{w}_k$&The dedicated precoding vector to user $k$ or group $k$\\
			\hline
			$z_k$& The noise at receiver $k$\\
			\hline
			$t$& Time index \\
			\hline
		\end{tabular}
		\vspace{0.2cm}
		\caption{\label{comparison1} Summary of the System Model Parameters  }
	\end{center}
\end{table}

\begin{figure*}[t]
\vspace{-1.5cm}
\hspace{0.2cm}
\begin{tabular}[t]{c}
\begin{minipage}{17 cm}
 \begin{center}
\hspace{-1.5cm}\includegraphics[scale=0.65]{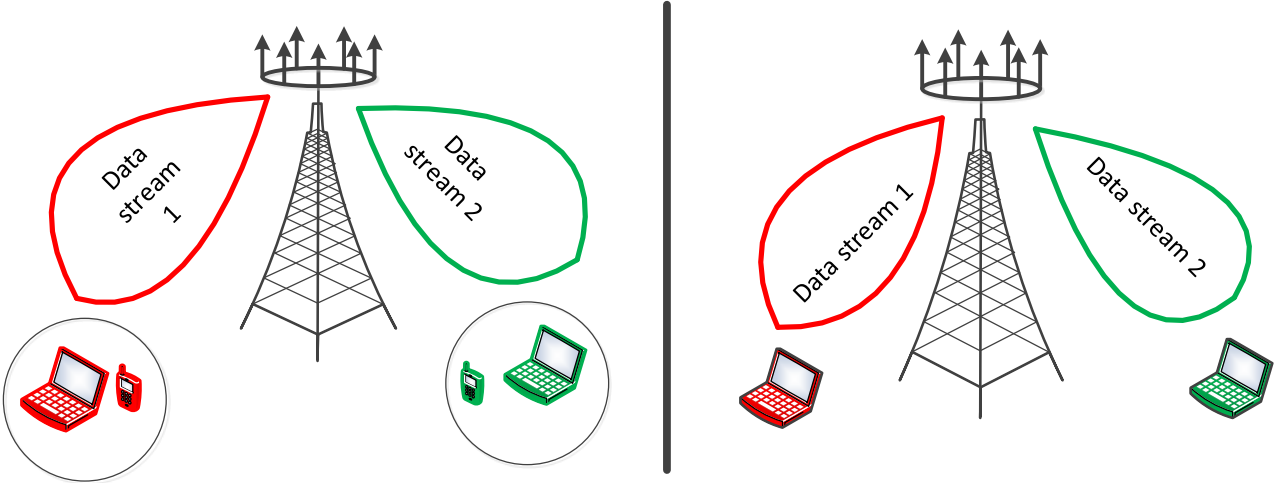}
\vspace{-7.5cm}
\caption{\label{UNICAST} System model for multicast (left) and unicast (right)}

\end{center}
\end{minipage}\\
\end{tabular}
\end{figure*}

\subsection{Classification Methodology}
\label{Classification}
\begin{figure}[h]
 	\vspace{-2cm}
 	\hspace{-2.0cm} \includegraphics[scale=0.95]{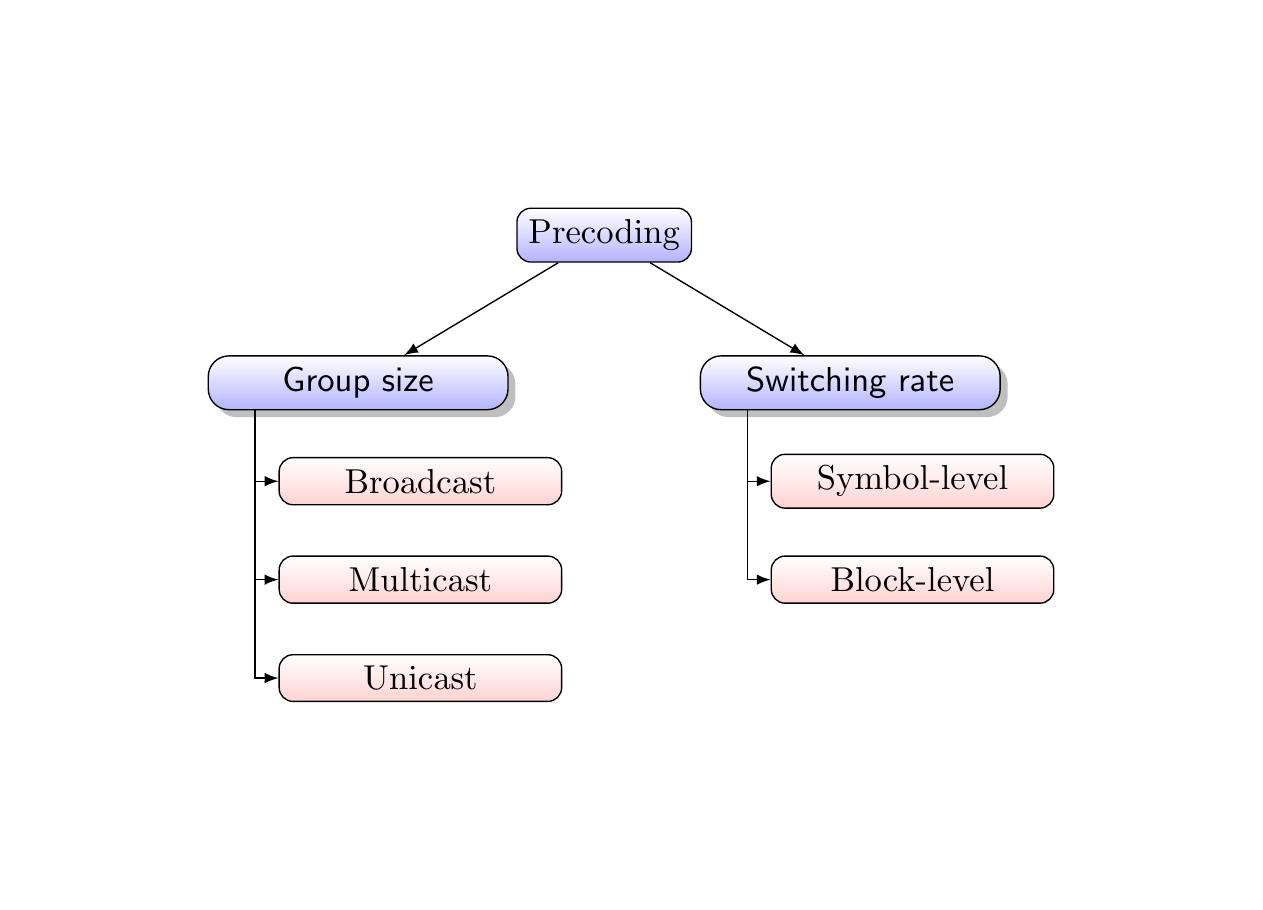}
 	\vspace{-2.5cm}
 	\caption{\label{tree2} Precoding Classification}
 \end{figure}

The adopted classification methodology is based on the tree of Fig. \ref{tree2}. 
   
\subsubsection{Block- vs Symbol-level precoding} The first classification axis is based on the switching rate of the precoding. \textbf{Block-level precoding} refers to techniques which apply precoding over symbol blocks. As a result, these techniques can use as side knowledge the channel matrix $\mathbf{H}$, which includes estimates of the channel coefficients for all antenna-user pairs. In this case, precoding refers to designing the covariance matrix of the output signal vector $E_t[\mathbf{x}\mathbf{x}^H]$. \textbf{Symbol-level precoding} refers to techniques where precoding is applied according to the baudrate. As a result, the techniques can use as side knowledge both the channel matrix $\mathbf{H}$ and the input symbol vector $\mathbf{s}_k[t]$. In this case, precoding refers to designing the actual output signal vector $\mathbf{x}[t]$.

\begin{figure*}[t]
	\vspace{-1.5cm}
	\hspace{0.2cm}
	\begin{tabular}[t]{c}
		\begin{minipage}{17 cm}
			\hspace{-4.5cm}\includegraphics[scale=0.35]{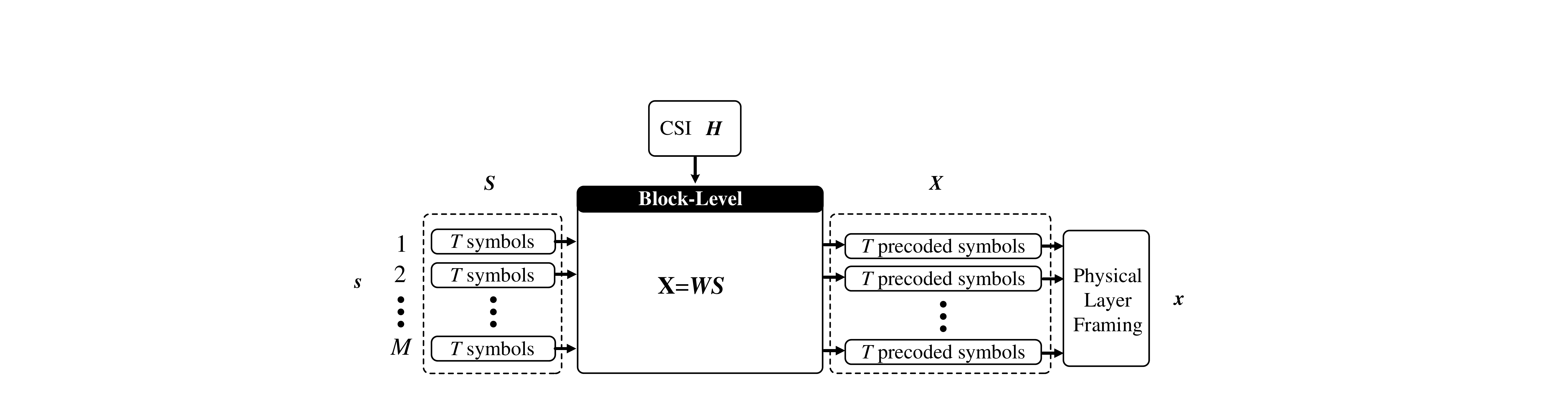}
			\vspace{-1.0cm}
			\caption{\label{Frame}  Schematic Diagram for Block-level Precoding transmitter in which the precoder changes with only CSI. }
			\vspace{-1cm}
			\hspace{-3cm}\includegraphics[scale=0.33]{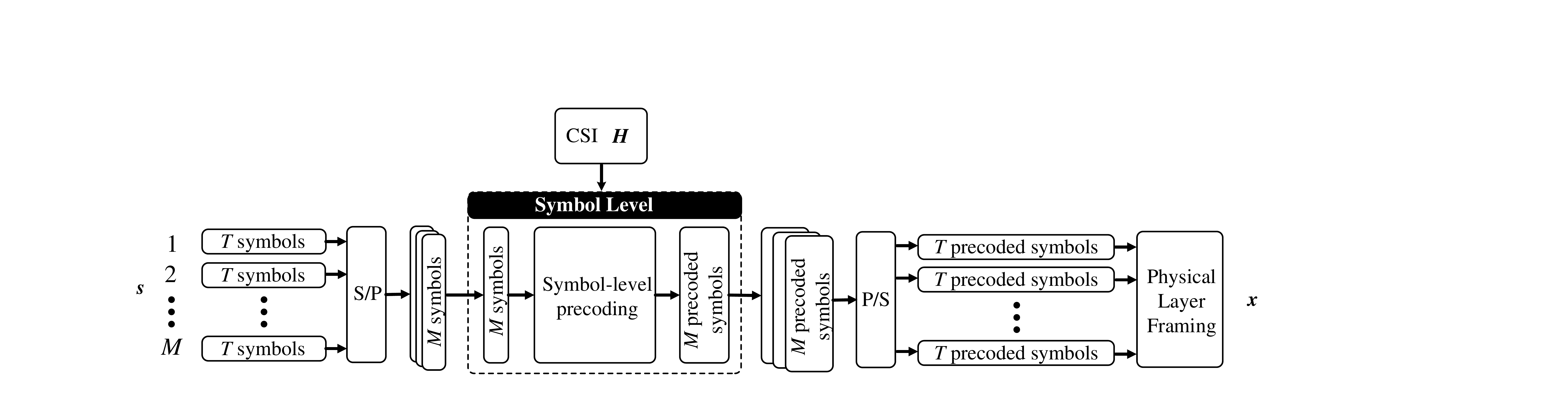}
			\vspace{-1.0cm}
			\caption{\label{Symbol} Schematic Diagram for Symbol-level Precoding transmitter in which the precoder changes with CSI and data symbols, where $\text{S/P}$ and $\text{P/S}$ denote the serial to parallel and parallel to serial parallel respectively. $\text{S/P}$ operation allows to design the prosssing at symbol-level.  }
				
		\end{minipage}\\
	\end{tabular}
\end{figure*}

\subsubsection{Uni-/Multi-/Broad-cast Precoding} The second classification axis is based on the number of targeted users per symbol stream. \textbf{Unicast precoding} refers to cases where each symbol stream is destined to a single intended user, i.e. $M=K$. \textbf{Broadcast precoding} (a.k.a. PHY-layer Multicasting) refers to cases where a single symbol stream is destined to all users, i.e. $M=1$. \textbf{Multicast precoding} refers to cases where each of $M$ symbol streams is destined to $M$ groups of $G$ intended users per group.

\subsubsection{Targeted Performance Metric} The precoder design techniques can be also differentiated based on the performance metric that they aim to optimize. The two main metrics in the literature are transmitted \textbf{Power} and \textbf{Quality of Service} (SNR, rate etc). The usual approach is to optimize one metric while using the other as a constraint, e.g. power minimization under QoS constraints, QoS maximization under power constraints\footnote{Often the two problems are dual and the power minimization is used as a stepping stone for solving the QoS maximization problem.}. 

\section{Preliminaries}
\label{Preliminaries}
\subsection{Power Metrics}
In this section, we formally define the basic power metrics that are going to be used in the remainder of this paper. By focusing on a specific symbol slot $t$ and a specific antenna $n$, the \textit{instantaneous per-antenna power} is defined as $P_n=|x_n[t]|^2$. Similarly, by focusing on a specific symbol slot $t$ and all $N$ antennas, the \textit{instantaneous sum power} is defined as $P=\|\mathbf{x}[t]\|^2=\mathrm{trace}(\mathbf{x}[t]\mathbf{x}^H[t])$. By averaging across multiple symbol slots and considering all antennas, the \textit{average sum power} is defined as $\bar P=E_t\big[\|\mathbf{x}[t]\|^2\big]=\mathrm{trace}(E_t\big[\mathbf{x}[t]\mathbf{x}^H[t]\big])$. Finally by averaging across multiple symbol slots and considering a single antenna $n$, the \textit{average per-antenna power} is defined as $\bar P_n=E_t\big[|x_n[t]|^2\big]=[E_t[\mathbf{x}[t]\mathbf{x}^H[t]]_{nn}$. 

One might wonder why we need so many different metrics. The answer is that each power metric serves a different purpose and can help address various implementation constraints or practical impairments. For example, average power provides an estimation of the long-term energy requirements, while instantaneous is a more detailed characterization, allowing to detect power spikes which could have unwanted side-effects. These side-effects include entering into the non-linear region of an amplifier or exceeding its maximum capability. Furthermore, the per-antenna power metrics are meant to enable the investigation of each RF chain individually. More specifically, one could check how the power is distributed across the multiple antennas, since each RF chain usually has its own amplifier with individual impairments and limitations.

\subsection{$\mathrm{QoS}$ metrics (SNR, rate)}
In this section, we formally define the basic $\mathrm{QoS}$ metrics that are going to be used in the remainder of this paper. A basic $\mathrm{QoS}$ metric is the Signal to Interference and Noise Ratio (SINR), which enables us to characterize or optimize a ratio of desired to undesired power levels. However, an even more meaningful metric for communication systems is the rate. The dependence of rate to the SINR greatly depends on the employed input symbol distribution. The vast majority of approaches in the area of block-level precoding have used Gaussian inputs as a way of allowing the rate to scale logarithmically with the SINR\footnote{It is worth mentioning some notable exceptions (\cite{Zhi_Ding_TWC}, \cite{Lozano_Mercury_TIT}-\cite{Perez_verdu_TIT})}. However, in practical systems uniform discrete constellations (modulations) are commonly used in conjunction with adaptive modulation based on SINR thresholds to allow the rate scaling. This consideration complicates the rate calculation, because each symbol block might use a different modulation whose performance has to be studied separately. As we will see in section \ref{Sub:Sec: SLP}, the vast majority of symbol-level techniques have adopted the latter mode, since the detection regions of the discrete modulations can be more easily modeled.  

\subsection{Block-level Unicast Precoding}
In this section, we briefly summarize some preliminaries on block-level unicast precoding, which is the most well-understood class in the literature. In this class, we could include dirty paper coding (DPC), which is an optimal non-linear technique based on known interference pre-cancellation which has been shown to achieve the MIMO downlink capacity~\cite{Costa1983,Weingarten2006}. Tomlinson-Harashima precoding (THP), which is a suboptimal implementation of DPC~\cite{Tomlinson1971}, could also be considered in the class of block-level precoding. Nevertheless, hereafter the focus is on linear block-level precoding approaches, characterized by a lower complexity, thus being more suitable for practical implementations.

In this framework, considering a block of $T$ symbol vectors to be conveyed to the users, modeled by an $M \times T$ matrix $\mathbf{S}$, the corresponding block representing the output signals can be written as:
\begin{eqnarray}\mathbf{X} = \mathbf{W}\mathbf{S}.\end{eqnarray}
The $N \times M$ matrix $\mathbf{W}$ is the precoding matrix, applied to the entire information block $\mathbf{S}$. The precoding matrix can be written as $\mathbf{W}=[\mathbf{w}_1 \ldots \mathbf{w}_M]$, each column represents a precoding vector for the corresponding user.
From this formalization, it is clear how the problem of block-level unicast precoding can be reduced to the problem of designing the precoding matrix  $\mathbf{W}$, using the knowledge of the channel $\mathbf{H}$, in order to mitigate the interference.
To this aim, the literature provides some closed-form as well as some solutions based on numerical optimization problems. 

The most relevant closed-form solutions are zero-forcing (ZF) precoding~\cite{Yoo2005,Yoo2006} and minimum mean square error (MMSE) precoding~\cite{Bjornson_SPM,Joham_2005,Peel2005,Yang_1997}. ZF is one of the simplest suboptimal techniques, which decouples the multi-user channel into parallel single-user channel, thus canceling out the multi-user interference. To this aim, the ZF precoding matrix can be calculated as the pseudo-inverse of the channel matrix, as $\mathbf{W} = \mathbf{H}^H(\mathbf{H}\mathbf{H}^H)^{-1}$. The ability of ZF precoding to cancel out the interference, makes it more appealing for the high SNR regime. However, since ZF does not take into account the effect of noise, it does not perform well in the low SNR regime (noise limited regime). MMSE precoding, on the other hand, takes into account both the interference and the noise in order to improve the system performance also in the noise limited scenarios\cite{Bjornson_SPM}. The MMSE precoding matrix can be written as $\mathbf{W}=\mathbf{H}^H(\mathbf{H}\mathbf{H}^H + \alpha\mathbf{I})^{-1}$, with $\alpha$ being a regularization parameter inversely proportional to the SNR. Because of its expression, the MMSE precoder is also referred to as regularized ZF (R-ZF)~\cite{Bjornson_SPM,Nguyen_2008,Muharar_2013}. It is worth mentioning also maximum ratio transmission (MRT) precoding~\cite{Lo1999}, aiming at maximizing the received SNR, which however is a suitable technique only in the noise limited regime, where the multi-user interference can be neglected.

The above mentioned closed-form solutions for precoding are effective and easy to implement. However, they do not allow to optimize the system with respect to specific objectives, or respecting specific constraints. In this regard, a number of optimization-based precoding techniques have been devised, so to enhance the flexibility of the transmitter. The literature on block-level precoding includes different optimization strategies for the precoding design. The optimal precoding strategy for the minimization of the transmitted average sum power, whilst guaranteeing some $\mathrm{QoS}$ targets at each user, was given in \cite{mats,Bengtsson1999}. For block level precoding, it can be shown that the average sum power is 
$\bar P = \sum_{j=1}^M\|\mathbf{w}_j\|^2$. Accordingly, the related optimization problem, which is optimally solved by semi-definite relaxation (SDR), can be written as follows:
\begin{equation}
\label{SPM_Unicast}
\begin{aligned}
\hspace{-0.1cm}\mathbf{W}(\mathbf{H},\mathbf{\gamma}) = \;
& \arg \;\underset{\mathbf{W}}{\min} 
\quad \sum_{j=1}^M\|\mathbf{w}_j\|^2 \\
& \text{s.t.} \quad \frac{|\mathbf{h}_j\mathbf{w}_j|^2}{\sum_{k\neq j,k=1}^M|\mathbf{h}_j\mathbf{w}_k|^2+\sigma_z^2 }\geq \gamma_j,\\
&\; j = 1, \dots, K, \\
\end{aligned}
\end{equation}
where the inputs are the channel matrix and a vector $\mathbf{\gamma}$ including the target SINR for the different users, and the output is the precoding matrix. 

Another relevant precoding strategy aims at maximizing the minimum SINR across the users, under sum power constraints (SPC). This approach increases the fairness of the system, thus it is known as \textit{max-min fair} optimization. The related optimization problem was solved in~\cite{boche} based on the principles of uplink/downlink duality, and can be written as:

\begin{equation}
\label{max_min_SPC_Unicast}
\begin{aligned}
\mathbf{W}(\mathbf{H},P) = \;
& \arg \;\underset{\mathbf{W}}{\max}\; \underset{j}{{\min}} 
\quad \frac{|\mathbf{h}_j\mathbf{w}_j|^2}{\sum_{k\neq j,k=1}^M|\mathbf{h}_j\mathbf{w}_k|^2+\sigma_z^2 } \\
& \text{s.t.} \quad \sum_{j=1}^M\|\mathbf{w}_j\|^2\leq P. \\
\end{aligned}
\end{equation}

Block-level precoding for unicast systems was extended in~\cite{Yu2007,Shen_2016} accounting for per-antenna power constraints. In particular, it is worth mentioning that the average per-antenna power can be written as $\bar P_n = \left[\sum_{j=1}^M  \mathbf w_j\mathbf w_j^H  \right]_{nn}$. Moreover, further developments have been done considering per-antenna-array power constraints~\cite{Dartmann2013} and non-linear power constraints~\cite{Zheng2012}.
 
Unicast multiuser MIMO techniques have been proposed to utilize the spatial multiplexing gains of MIMO for different network capabilities such as multicell MIMO \cite{Dahrouj_2010}, cognitive radio \cite{Nguyen_2016_cognitive}, physical layer security \cite{Secure_SWIPT_2015,MU_MIMO_security}, simultaneous wireless information and power transfer \cite{Secure_SWIPT_2015,Xu_2014_SWIPT}, etc.   
\section{Block-level Multicast Precoding} \label{sec:block:level:pre}
A fundamental consideration of the multiuser unicast precoding is that independent data is addressed to each user. However, the new generation of multi-antenna communication standards has to adapt the physical layer design to the needs of the higher network layers. Examples of such cases include highly demanding applications (e.g. video broadcasting) that stretch the throughput limits  of multiuser broadband systems. In this direction, physical layer ($\mathrm{PHY}$) multicasting has the potential to efficiently address the nature of future traffic demand and  has become part of the new generation of communication standards. $\mathrm{PHY}$ multicasting is also relevant for the application of beamforming without changing the framing structure of standards (cf. \cite{Christopoulos2014TWCOM}).

\subsection{Multicast}

In the framework of block-level multicast precoding, we assume multiple interfering  groups of users. In each group, each user  receives a  stream of common data. However, independent symbols are addressed to different groups and inter-group interferences comes into play. A unified framework for physical layer multicasting to multiple co-channel groups, where independent sets of common data are transmitted to  groups of users by the multiple antennas, was given in \cite{Karipidis2005CAMSAP,Karipidis2007,Karipidis2008}. Therein, the  $\mathrm{QoS}$ and the fairness problems were formulated, proven NP-hard and solved for  the sum power constrained multicast multigroup case. The $\mathrm{QoS}$ problem, aiming at minimizing the average sum transmit power, has been solved resorting to SDR, and can be written as:

\begin{equation}
\label{SPM_Multicast}
\begin{aligned}
\mathbf{W}(\mathbf{H},\mathbf{\gamma}) = \;
& {\arg} \;\underset{\mathbf{W}}{\min} 
\quad \sum_{k=1}^M\|\mathbf{w}_k\|^2 \\
& \text{s.t.} \quad \frac{|\mathbf h_i\mathbf w_k|^2}{\sum_{{l\neq k }}|\mathbf h_i\mathbf w_l|^2+\sigma_i^2 }\geq \gamma_i, \\\; &\forall i \in\mathcal{G}_k, k, l\in\{1\dots M\}, \\
\end{aligned}
\end{equation}
where $\mathbf w_k\in \mathbb{C}^{N_t}$, and $\mathcal{G}_k$ denotes the $k$-th group of users. The notation $\sum_{l \neq k}$ states that aggregate interference from all co-channel groups is calculated.

The weighted max-min fair problem under sum power constraints (SPC) has been solved via bisection over the $\mathrm{QoS}$ problem, and can be written as:

\begin{eqnarray}\nonumber
\label{max-min_SPC_Multicast}
\mathbf{W}(\mathbf{H},P) =
& \arg\underset{t,\mathbf{W}}{\min} 
\quad t \\\nonumber
& \text{s.t.} \quad \frac{1}{\gamma_i}\frac{|\mathbf h_i\mathbf w_k |^2}{\sum_{{l\neq k }}|\mathbf h_i\mathbf w_l|^2+\sigma_i^2 }\geq t,\\\; &\forall i \in\mathcal{G}_k, k, l\in\{1\dots M\}, \\\nonumber
& \quad\sum_{k=1}^M\|\mathbf{w}_k\|^2\leq P,
\end{eqnarray}
where $\mathbf w_k\in \mathbb{C}^{N}$ and $t \in \mathbb{R}^{+}$.  Different service levels between the users can be acknowledged in this weighted  formulation. The problem receives as inputs the SPC $P$ and the target $\mathrm{SINR}$s vector $\mathbf \gamma = [\gamma_1,\gamma_2, \dots \gamma_{K}]$. Its goal is to maximize the slack variable $t$ while keeping all $\mathrm{SINR}$s above this value. Thus, it constitutes a max-min problem that guarantees fairness amongst users. Of particular interest is the case where the co-group users  share the same target i.e. $\gamma_i = \gamma_{k},\ \forall i \in\mathcal{G}_k, k\in\{1\dots G\}$.

The weighted max-min fair problem has been addressed also accounting for per-antenna power constraints (PACs). In the related optimization problem, analogous to \eqref{max-min_SPC_Multicast}, the PACs read as $\left[\sum_{k=1}^M  \mathbf w_k\mathbf w_k^H  \right]_{nn}  \leq P_n, \forall n\in \{1\dots N_{t}\}$. The weighted max-min fair problem with PACs has been solved through different approaches, as discussed hereafter.

\subsubsection{SDR-based solution}
The optimal multigroup multicast precoders when a maximum limit is imposed on the transmitted power of each antenna, have been derived in \cite{Christopoulos2014_ICC,Christopoulos2014_TSP}.   Therein,  a consolidated solution for the weighted max--min fair multigroup multicast beamforming problem under per-antenna constraints ($\mathrm{PAC}$s)   is  presented.  This framework is based on $\mathrm{SDR}$ and Gaussian randomization to solve the $\mathrm{QoS}$ problem and bisection to derive an accurate approximation of the non-convex $\max \min$ \textit{fair }formulation. However, as detailed  in \cite{Christopoulos2014_TSP}, the $\mathrm{PAC}$s are bound to increase the complexity of the optimization problem and reduce the accuracy of the approximation, especially as the number of transmit antennas is increasing. These observations necessitate the investigation of lower complexity, accurate approximations that can be applied on large-scale antenna arrays, constrained by practical, per-antenna power limitations.
\subsubsection{Successive Convex Approximation based solution}
 Inspired by the recent development of the feasible point pursuit ($\mathrm{FPP}$) successive convex approximation ($\mathrm{SCA}$) of non-convex quadratically constrained quadratic problems ($\mathrm{QCQP}$s), as developed in \cite{Sidiropoulos2015_SPL}, the work of \cite{Christopoulos2015_SPAWC} improved the $\max \min$ \textit{fair }  solutions of \cite{Christopoulos2014_TSP} in terms of computational complexity and convergence. The $\mathrm{FPP-SCA}$  tool  has been preferred over other existing approximations (for instance \cite{Sidiropoulos2015_SPL}) due to its guaranteed feasibility regardless of the initial state of the iterative optimization \cite{Sidiropoulos2015_SPL}. 
 
Apart from these two major approaches for solving multicast beamforming problems, an iterative technique recently appeared in literature \cite{Demir2016}. In this paper, the QoS problem was cast in a equivalent form and then an iterative method based on alternating minimization was developed for its solution. This approach does not rely on optimization toolboxes and exhibits significant reduced computational complexity compared to the two other approaches while it achieves in general better performance than the SDR approach and very close to the SCA one. Furthermore, this approach was extended to the hybrid analog-digital transceivers' case which have growing interest the last years due to the recent developments in mmWave and Massive MIMO systems. Further approaches that investigate the potential of multicast beamforming schemes in hybrid transceivers or in general large array systems can be found in \cite{Dai2016, Demir2016, Mehanna2013, Christopoulos2015_SPAWC}.  
 
\subsection{Broadcast}
Broadcast precoding can be seen as a special case of multicast, where we have a single group of users receiving the same data information. In this scenario, there is no interference since a single stream is sent to all users. In \cite{Sidiropoulos:2006}, the NP-hard broadcast precoding problem was accurately approximated by $\mathrm{SDR}$ and Gaussian randomization. The associated $\mathrm{QoS}$ problem can be written as:

\begin{equation}
\label{SPM_Broadcast}
\begin{aligned}
\mathbf{w}(\mathbf{H},\mathbf{\gamma}) = \;
& \arg \;\underset{\mathbf{w}}{\min} 
\quad \|\mathbf{w}\|^2 \\
& \text{s.t.} \quad \frac{|\mathbf h_i\mathbf w|^2}{\sigma_i^2 }\geq \gamma_i, \; \; j = 1, \dots, K, \\
\end{aligned}
\end{equation}
where $\mathbf w\in \mathbb{C}^{N_t}$ represents the precoding vector for the unique transmitted data stream.

In 5G wireless network, we expect a dramatic increase in services and applications \cite{Zander_2013}.  Employing an integrated framework of broadcast, multicast and unicast depending on the content of requested streams improves the efficiency of the wireless networks  \cite{Christopoulos_2015_scvt,Larsson_2016_broadcast,Tao_2016_yu,Wu_2015_Hyper}. For example,  using multicast solely, the rate of each link is limited by the  worst user wasting a considerable link margin available for delivering extra information. To deal with this inefficiency,  a multiuser MIMO system that enables a joint utilization of broadcast, unicast, and multicast is required. This can efficiently leverage the unused MIMO capability to send a broadcast stream or unicast streams concurrently with multicast ones, while ensuring no harm  to the achievable rate of multicasting. Therefore, the throughput and energy efficiency of the whole network can be improved significantly.  For more details about the application of block-level precoding, please look at Table \ref{comparison2}.
\\  
In the next section, we will discuss the classification based on switching rate.
\section{Symbol-level Unicast Precoding}\label{sec:slp:unicast}
As observed in the block-level precoder class in Section~\ref{sec:block:level:pre}, precoding at the transmitter is used to mitigate the interference among the users' data streams. As another approach, the data and channel information can be used to perform symbol-level precoding at the transmitter. Symbol-level precoding guarantees interference-free communication at the expense of higher switching rate of the precoder. In the literature, symbol-level precoding paradigm has been proposed in two different research avenues, namely, directional modulation, via analog symbol-level precoding, developed in antenna and propagation domain and digital symbol-level precoding for constructive interference { developed in signal processing and wireless communications}. The solutions of both of these approaches are developed under the same context of channel and data dependent precoding, they originate from different areas and function under different system level models though. Thus, each one of them shares different advantages and disadvantages and comes with a different number of challenges that must be overcome towards the implementation of efficient transceiver solutions.

A transceiver based on the directional modulation concept consists of only a single RF chain { which is fed by a local RF oscillator. The RF chain} drives a network of phase shifters and variable gain amplifiers. In this technology, the antennas excitation weights change in the analog domain on a symbol basis, to create the desired phase and amplitude at the receiver side{instead of generating the symbols at transmitter and sending them}. While a single RF chain transceiver is highly desirable due to its simplistic structure and power consumption, there are several limitations regarding implementation difficulties and the lack of a strong algorithmic framework that need further study in the directional modulation field. 
\begin{figure*}[t]
	\vspace{-0.5cm}
	\hspace{0.2cm}
	\begin{tabular}[t]{c}
		\begin{minipage}{17 cm}
		 \includegraphics[scale=0.8]{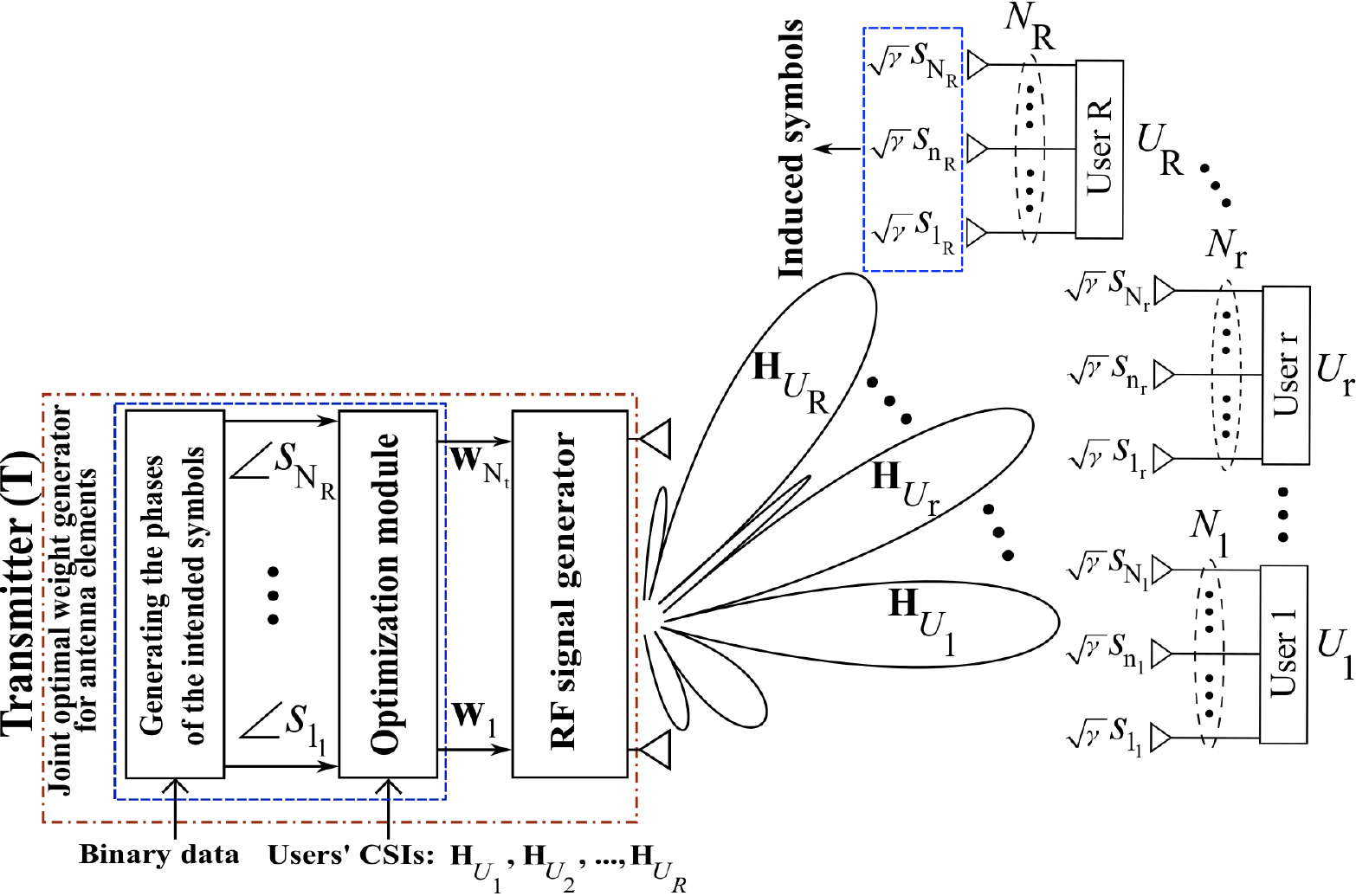}
		 \caption{\label{DM_FIg_1} Generic {structure of a directional modulation transmitter}}
		 \vspace{1cm}\begin{center}
		 \includegraphics[scale=0.8]{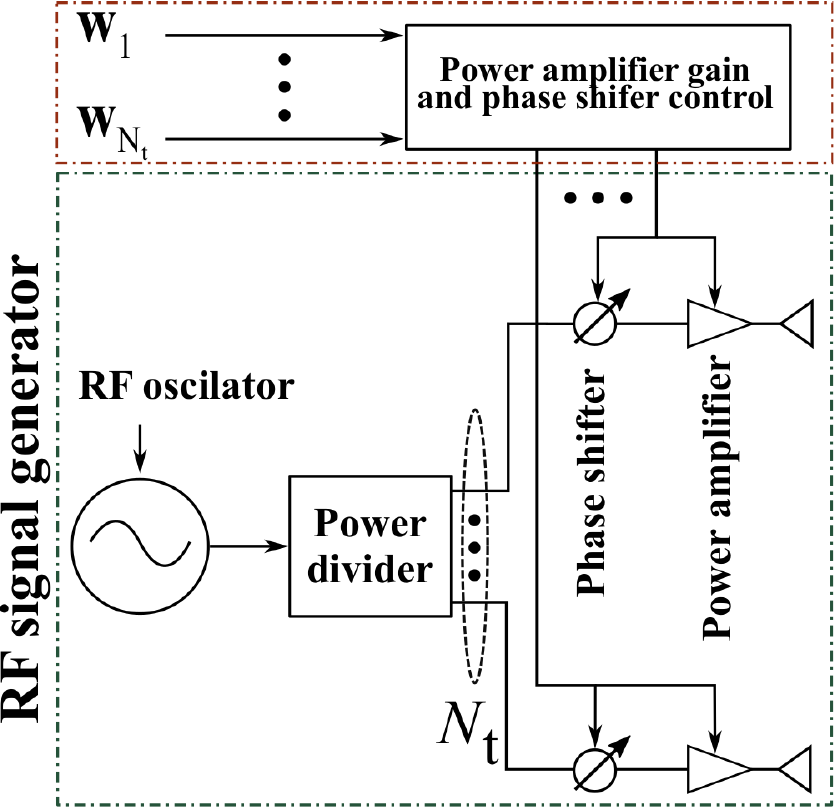}
		 \caption{\label{DM_FIg_2} Detailed schematic diagram for a directional modulation { transmitter} (analog symbol-level precoding)}
			
				\end{center}
		\end{minipage}\\
	\end{tabular}
\end{figure*}

On the other hand, the digital symbol-level precoding for constructive interference uses digital precoding for signal design at the transmitter in order to create constructive interference at the receiver. The digital precoding happens before feeding the signal to the antenna array. {Symbol-level precoding developed in the signal processing and wireless communications domain and the related techniques are more studied from an algorithmic point of view compared to the directional modulation based ones. On the contrary, they require a full digital transceiver, and thus there is difficulty in applying them in large antenna array systems.} 

In the following, a detailed description of directional modulation and digital symbol-level precoding { are presented} to show the differences and the similarities of the both schemes.
\subsection{Symbol-Level Precoding for Directional Modulation}

Directional modulation is an approach in which the users' channels and symbols are used to design the phase and amplitude of each antenna on a symbol basis such that multiple interference-free symbols can be communicated with the receiver(s). After adjusting the array weights, the emitted radio frequency (RF) signals from the array are modulated while passing through the fading channel. This is different from block-level precoding in which the transmitter generates the symbols and sends them after precoding~\cite{Lai-U:2004,Spencer:2004}. The benefit of directional modulation is that the precoder is designed such that the receivers antennas can directly recover the symbols without CSI and equalization.  In Fig. \ref{DM_FIg_1}-\ref{DM_FIg_2}, a transmitter architecture for directional modulation is depicted.

Recently, there has been growing research interest on the directional modulation technology. Array switching approach at the symbol rate is used in~\cite{Baghdady:1990,Daly:patt:2010,zhu:4D:array:2014} to induce the desired symbols at the receiver side. Specifically, the work of~\cite{Baghdady:1990} uses an antenna array with a specific fixed delay in each RF chain to create the desired symbols by properly switching the antennas. The authors in~\cite{Daly:patt:2010} use an array where each element can switch to broadside pattern\footnote{Maximum radiation of an array directed normal to the axis of the array.}, endfire pattern\footnote{Additional maxima radiation directed along the axis.}, or off status to create the desired symbols in a specific direction. The authors perform an extensive exhaustive search to find the best combination among the antenna patterns. In the work of~\cite{zhu:4D:array:2014}, the elements of the array are switched to directionally modulate the $\mathit{B-PSK}$ constellation. 

In another category, parasitic antenna is used to create the desired amplitude and phase in the far field by near field interactions between a driven antenna element and multiple reflectors~\cite{Babakhani:2008,Babakhani:2009,Lavaei:2010}. As pioneers in this approach,~\cite{Babakhani:2008,Babakhani:2009} use transistor switches or varactor diodes to change the reflector length or its capacitive load, respectively, when the channel is line of sight (LoS). This approach creates a specific symbol in the far field of the antenna towards the desired directions while randomizing the symbols in other directions due to the antenna pattern change. In connection with~\cite{Babakhani:2008},~\cite{Lavaei:2010} studies the far field area coverage of a parasitic antenna and shows that it is a convex region. 

The authors of~\cite{Daly:2009} suggest using a phased array at the transmitter, and employ the genetic algorithm to derive the phase values of a phased array in order to create symbols in a specific direction. The technique of~\cite{Daly:2009} is implemented in~\cite{Daly:2010} using a four element microstrip patch array where the genetic algorithm is used to derive the array phase in order to directionally modulated the symbols based $\mathit{Q}$-PSK modulation. The authors of~\cite{Daly:2011} propose an iterative nonlinear optimization approach to design the array weights which minimizes the distance between the desired and the directly modulated symbols in a specific direction. In another paradigm, the authors of~\cite{Guo_2016}, divide the interference into static and dynamic parts and use genetic algorithms to design the array weights to directionally modulate the symbols.

In~\cite{Tao:2011}, baseband in-phase and quadrature-phase signals are separately used to excite two different antennas so that symbols are correctly transmitted only in a specific direction and scrambled in other directions. In another paradigm,~\cite{Valliappan:2013} uses random and optimized codebook selection, where the optimized selection suppresses large antenna side lobes, in order to improve the security in a millimeter-wave large uniform linear antenna array system. The authors of~\cite{Ding:2014} derive optimal array weights to get a specific bit error rate (BER) for $\mathit{Q}$-PSK modulation in the desired and undesired directions. The Fourier transform is used in~\cite{Yuan:2013:fourier,Yuan:2014:con} to create the optimal constellation pattern for $\mathit{Q}$-PSK directional modulation. The work of~\cite{Yuan:2013:fourier} uses Fourier transform to create the optimal constellation pattern for $\mathit{Q}$-PSK directional modulation, while~\cite{Yuan:2014:con} uses Fourier transforms along with an iterative approach for $\mathit{Q}$-PSK directional modulation and constraining the far field radiation patterns. The effect of array structure on the directional modulation performance is investigated in~\cite{shi:enhance:2013}. The authors have shown that by increasing the space between the antennas of a two element array the symbol error rate can be improved for 8-PSK modulation. As an overview,~\cite{ding:met:2014} categorizes the directional modulation systems for $\mathit{Q}$PSK modulation and discusses the proper metrics such as bit error rate for evaluating the performance of such systems. To overcome imperfect measurements, the authors of~\cite{Hu:robust:2016} propose a robust design for directional modulation in the presence of uncertainty in the estimated direction angle. The authors use minimum mean square error to minimize the distortion of the constellation points along the desired direction which improves the bit error rate performance. 

In~\cite{Yuan:2014:con,Yuan:2014:patt:sep,Yuan:2015,Ding:2015:MIMO} directional modulation is employed along with noise injection. The authors of~\cite{Yuan:2014:con,Yuan:2014:patt:sep,ding:vec:2014} utilize an orthogonal vector approach to derive the array weights in order to directly modulate the data and inject the artificial noise in the direction of the eavesdropper. The work of~\cite{Yuan:2015:orth:mul:bm} is extended to retroactive arrays\footnote{A retroactive antenna can retransmit a reference signal back along the path which it was incident despite the presence of spatial and/or temporal variations in the propagation path.} in~\cite{Yuan:2015} for a multi-path environment. An algorithm including exhaustive search is used in~\cite{Hongzhe:2014} to adjust two-bit phase shifters for directly modulating information. The work of~\cite{ding:vec:2014} introduces vector representations to link the vector paths and constellations. This helps figuring out the transmitter characteristics and the necessary and sufficient condition for directionally modulating symbols. It is shown that the directional modulation can be realized by adjusting the gain of the beamforming network. 

The directional modulation concept is also extended to directionally modulate symbols to more than one destination. In~\cite{Ding:2015:MIMO}, the singular value decomposition (SVD) is used to directionally modulate symbols in a two user system. The authors of~\cite{Yuan:2015:orth:mul:bm} derive the array weights to create two orthogonal far field patterns to directionally modulate two symbols to two different locations and~\cite{Hafez:2015} uses least-norm to derive the array weights and directionally modulate symbols towards multiple destinations in a multi-user multi-input multi-output (MIMO) system. Later,~\cite{Hafez_2016} considers using ZF precoder to directionally modulate symbols and provide security for multiple single-antenna legitimate receivers in the presence of multiple single-antenna eavesdroppers. As a new approach, a synthesis free directional modulation system is proposed in~\cite{Ding_2016} to securely communicate information without estimating the target direction. 


The works of~\cite{Kalantari:DM:ICASSP:2016,Kalantari:JSTSP} design the optimal symbol-level precoder for a security enhancing directional modulation transmitter in a MIMO fading channel to communicate with arbitrary number of users and symbol streams. In addition, the authors derive the necessary condition for the existence of the precoder. The power and SNR minimization precoder design problems are simplified into a linearly-constrained quadratic programming problem. For faster design, an iterative approach as well as non-negative least squares formulation are proposed. 
\subsection{Symbol-level Precoding for Constructive Interference}
\label{Sub:Sec: SLP}
 The interference among the multiuser spatial streams
leads to a deviation of the received symbols outside of their detection region.
Block-level precoding treats the interference as harmful factor that should be mitigated \cite{Spencer:2004,Yoo2005,Yoo2006,Yu2007,mats,Bengtsson1999,Dartmann2013}. In this situation (see Fig. \ref{Interference_BLP}), the precoding cannot tackle the interference at each symbol and tries to mitigate the interference along the whole frame using only the knowledge of CSI, which manages to reduce the average amount of interference  along the frame. 
 \begin{figure}[h]
 \begin{center}
  \includegraphics[scale=0.5]{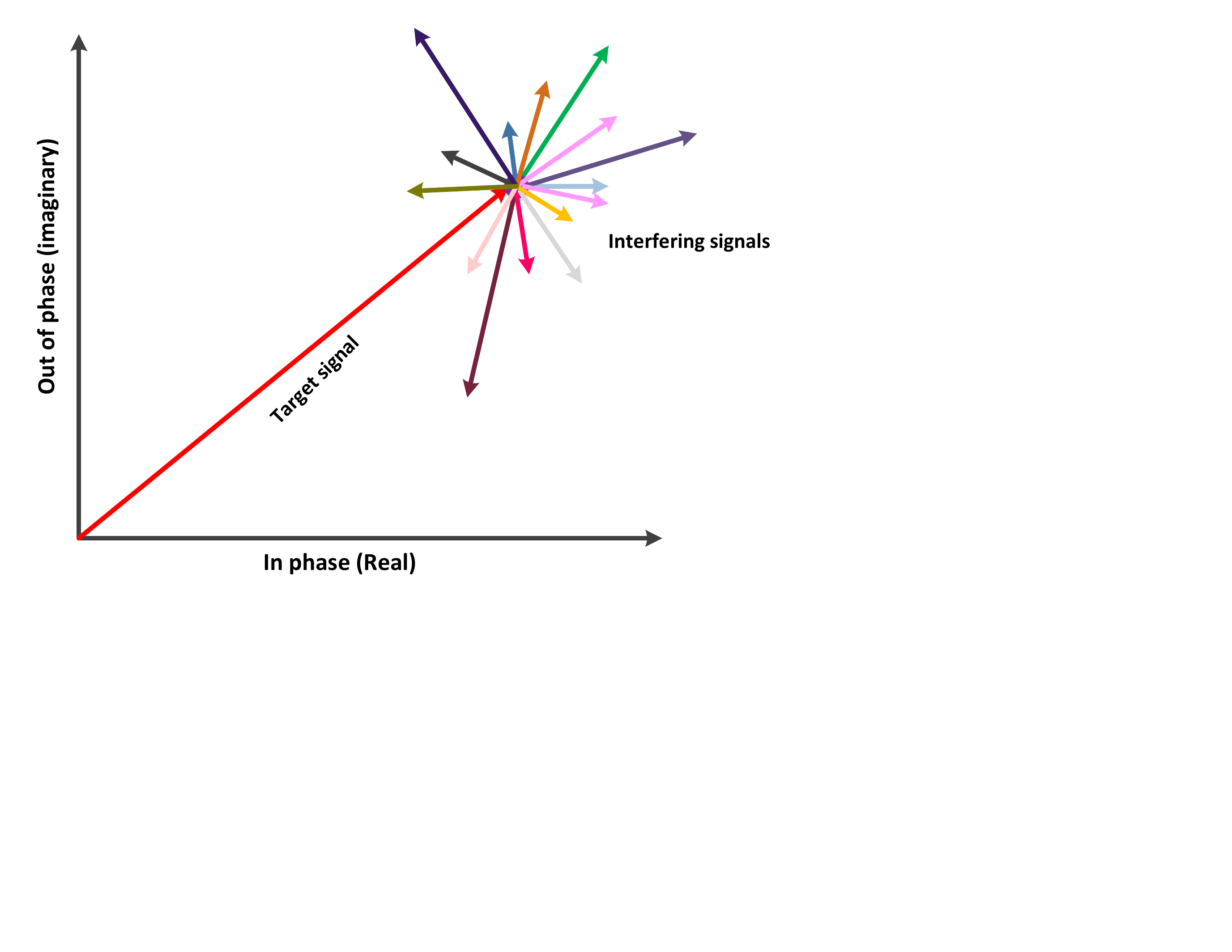}
  \vspace{-4.2cm}
  \caption{\label{Interference_BLP}Interference in Block-level Precoding. Interference can only be managed along whole frame.}
  \end{center}
 \end{figure}

\begin{figure}[h]
 \begin{center}
  \includegraphics[scale=0.5]{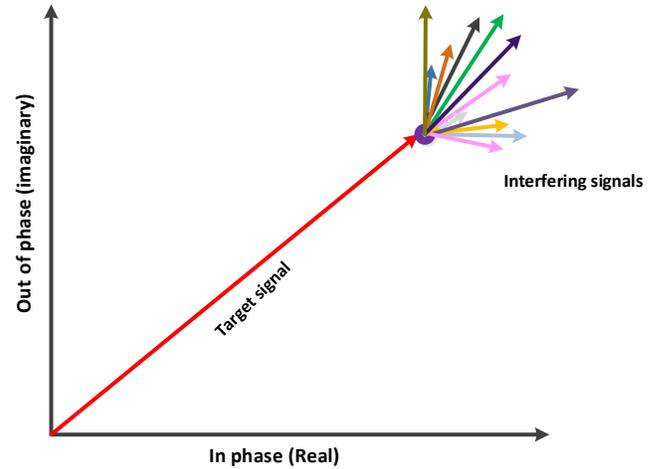}
  \vspace{-4.2cm}
  \caption{\label{Interference_BLP}Interference controlled on symbol by symbol basis to guarantee that the interference is constructive in symbol-level Precoding.   }
  \end{center}
 \end{figure}
 
During the past years several symbol-level processing techniques has been utilized in the multiuser MISO context \cite{Masouros2009,Masouros2011,Masouros_2012,Masouros_THP,Masouros_khan,maha_ISIT,maha_crowncom,maha_TSP,maha_SPAWC,maha_globecom,maha_twc,Spano2016Globecom,Alodeh_TWC2,Alodeh2016DSP,Spano2016TSP,SLP_Korean_2014,Masouros:2015,SLP_Korean_2016,Ntougias_2016,Amadori_2017}.  
A similar concept to the  symbol-level precoding is the so-called constant envelop precoding that appeared recently in the literature \cite{Mohammed2012,Pan2014,Mohammed2013,Chen2014,Mohammed2013b,Mukherjee_2015,Zhanq2016,SZhanq2016,SZhanq2016b}. In these techniques, constant modulus constraints are set to the complex baseband signal of each transmit antenna which is designed such that the difference between the noise free received signal at the receiver(s) and the desired symbol information is minimized in a least squares sense. The constant envelop based techniques exhibit low peak-to-average power ratio (PARP) and their concept presents similar advantages to the one of the directional modulation based transceivers, since ideally they can be also implemented in transceivers of a single RF chain that drives a phase shift network. On the contrary, the involved optimization problems are non-convex due to the constant modulus requirements and thus, they are hard to solve, they support restricted set of constellation points and they treat the interference like the block-level solutions, that is as a harmful component. For now and on we will focus our discussion on the symbol level precoding works.

\begin{figure*}
\centering
\begin{minipage}{\textwidth}
\includegraphics[width=1.1\textwidth]{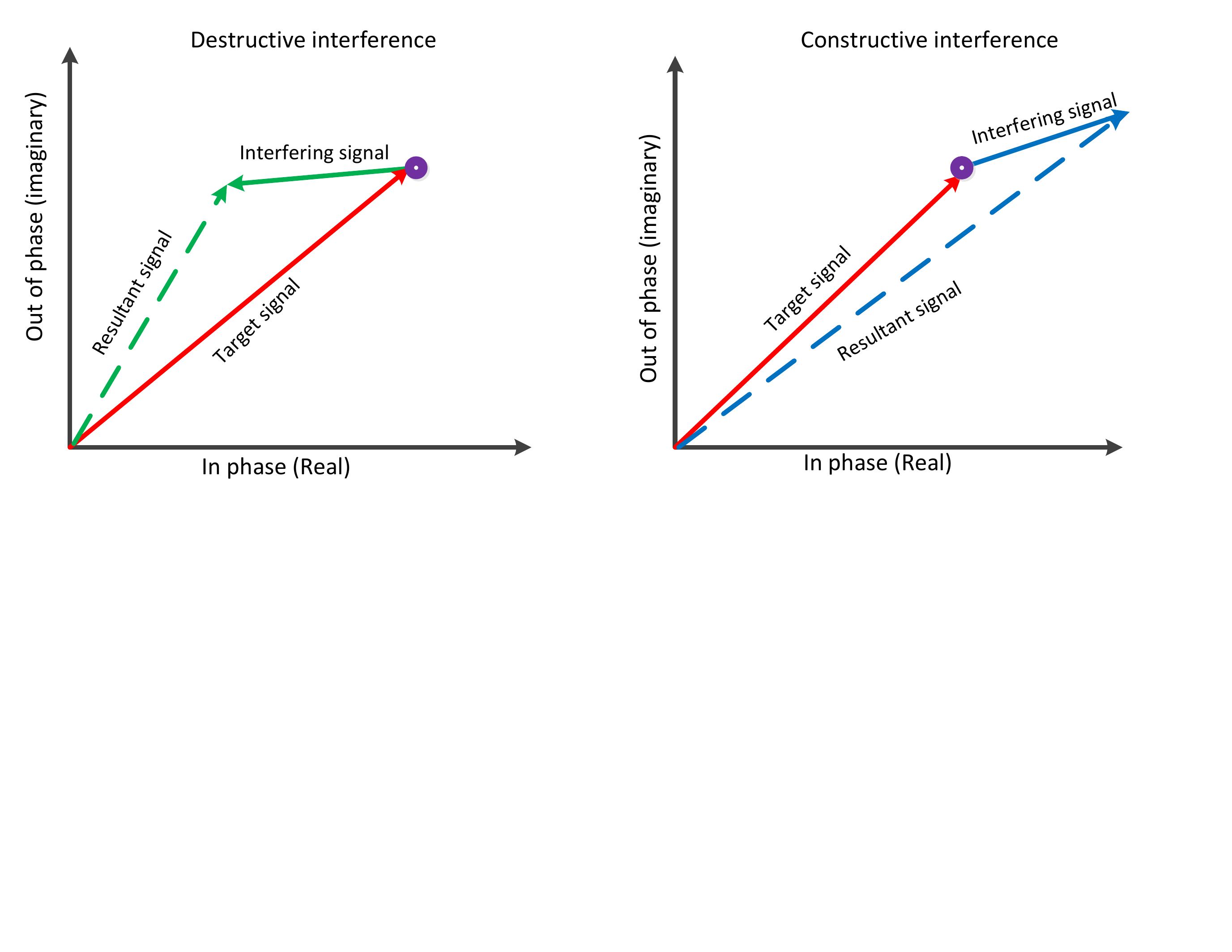}
\vspace{-7.5cm}
\caption{\label{classification}The first quadrant of $\mathit{Q}$-PSK. The Interference can be destructive as the figure in the left or constructive as the figure in the right.}
\end{minipage}
\end{figure*}

The interference can
be classified into constructive or destructive based on whether it facilitates or deteriorates the correct detection of the received symbol.  A detailed classification of interference is thoroughly discussed in for $\mathit{B}$-PSK and $\mathit{Q}$-PSK in\cite{Masouros2009} and for $\mathit{M}$-PSK in \cite{maha_TSP}. The constructive interference pushes the detected constellation point deeper into detection region. Fig. \ref{classification} illustrates the two scenario when the interference is destructive  and when it is constructive for $\mathit{Q}$-PSK modulation.

To classify the multiuser interference, both the data information and the CSI should be available at the transmitter.  the unit-power
 created interference from the $k^{th}$ data stream on the $j^{th}$ user can be formulated as:
\vspace{-0.1cm}
\begin{equation}
\psi_{jk}=\frac{\mathbf{h}^\dagger_{j}\mathbf{w}_k}{\|\mathbf{h}_{j}\|\|\mathbf{w}_k\|}.
\end{equation}
An $\mathit{M}$PSK modulated symbol $d_k$, is said to receive constructive
interference from another simultaneously transmitted symbol $d_j$ which is
associated with $\mathbf{w}_j$ if and only if the following inequalities hold   
\begin{equation}\nonumber
\label{one}
\angle{s_j}-\frac{\pi}{M}\leq \arctan\Bigg(\frac{\mathcal{I}\{\psi_{jk}s_{k}\}}{\mathcal{R}\{\psi_{jk}s_{k}\}}\Bigg)\leq \angle{s_j}+\frac{\pi}{M},
\end{equation}
\begin{equation}\nonumber
\label{two}
\mathcal{R}\{{s_k}\}.\mathcal{R}\{\psi_{jk}
s_{j}\}>0, \mathcal{I}\{{s_k}\}.\mathcal{I}\{\psi_{jk}s_{j}\}>0.\\
\end{equation}
This was proved in details \cite{maha_TSP}. One of the interesting characteristics of the constructive  interference between two streams is its mutuality. In more details, if the stream $\mathbf{w}_js_j$ constructively interferes with $\mathbf{w}_ks_k$ (i.e. pushes $s_k$ deeper in its detection region), then
the interference from transmitting the stream $\mathbf{w}_ks_k$ 
is constructive to $s_j$ \cite{maha_TSP}.

For constructively interfering symbols, the value of the received signal can be bounded as
\begin{eqnarray}\nonumber
\vspace{-0.1cm}
\label{Rxs}
\sqrt{p}_j\|\mathbf{h}_j\|\overset{(a)}{\leq} |y_j|\overset{(b)}{\leq} \|\mathbf{h}_{j}\|\big(\sqrt{p}_j+\displaystyle\sum^{K}_{\forall
k,k\neq j}\sqrt{p}_k|\psi_{jk}|\big).
\vspace{-0.5cm}
\end{eqnarray}
The inequality (a) holds when all simultaneous users are orthogonal (i.e. $\psi_{jk}=0$), while (b) holds when all created interference
is aligned with the transmitted symbol as $\angle d_k=\angle\psi_{jk} d_j$
 and $\psi_{jk}=0$, $\angle d_k=\angle\psi_{jk} d_j$.  
The previous inequality indicates that in the case of constructive
interference, having fully correlated signals is beneficial as they contribute
to the received signal power.  For a generic symbol-level precoding, the previous inequality can be 
\begin{eqnarray}\nonumber
0\overset{(a)}{\leq} |y_j|\overset{(b)}{\leq} \|\mathbf{h}_{j}\|\big(\sqrt{p}_j+\displaystyle\sum^{K}_{\forall
k,k\neq j}\sqrt{p}_k|\psi_{jk}|\big).
\vspace{-0.5cm}
\end{eqnarray}
In comparison to block-level precoding techniques, the previous inequality can be reformulated as
\begin{eqnarray}\nonumber
\label{rxs}
\vspace{-0.6cm}
0\overset{(a)}{\leq} |y_j| \overset{(b)}{\leq} \sqrt{p_j}\|\mathbf{h}_j\|.
\end{eqnarray}
The worst case scenario can occur when all users are co-linear, that is when $\psi_{jk}\rightarrow
1$. The channel cannot
be inverted and thus the interference cannot be mitigated. The optimal scenario takes place
when all users have physically orthogonal channels which entails no multiuser interference. Therefore, utilizing CSI and DI leads to higher performance in comparison to employing conventional techniques.
\subsubsection{Techniques}
 The difference between the block-level and symbol-level precoding techniques is illustrated in Fig. \ref{Frame}-\ref{Symbol}.
Fig. \ref{Frame} shows how the block-level precoding depends only on the CSI information to optimize $\mathbf{W}$ that
carry the data symbols $\mathbf{s}$ and without any design dependency
between them. In contrary, symbol-level precoding as illuastrated in Fig. \ref{Symbol} depends on both CSI and the data symbol combinations to optimize the precoding matrix $\mathbf{W}$ and the output vector $\mathbf{x}$. The optimal design for symbol-level precoding depends on how to define the optimization problem and more importantly how to define the constructive interference constraints. In~\cite{maha_ISIT,maha_TSP,maha_SPAWC,maha_crowncom,maha_globecom,maha_twc,Alodeh_TWC2}, the optimal precoding strategy for the minimization of the total transmit power, whilst guaranteeing QoS targets at each user, was given. For any generic modulation, the related optimization problem can be written as follows:
\begin{eqnarray}\nonumber
\mathbf{w}_k(\mathbf{s},\boldsymbol\gamma, \mathbf{H})&=&\arg\underset{\mathbf{w}_k}{\min}\|\sum^K_{k=1}\mathbf{w}_ks_k\|^2\\\nonumber
&s.t.& |\mathbf{h}_j\sum^K_{k=1}\mathbf{w}_k s_k|^2\geq \gamma_j\sigma^2, \forall j\in K\\
&\quad&\angle(\mathbf{h}_j\sum^K_{k=1}\mathbf{w}_k s_k)=\angle s_j,
\end{eqnarray}
by using $\mathbf{x}=\sum^K_{k=1}\mathbf{w}_k s_k$, the previous optimization can be formulated as:
\begin{eqnarray}\nonumber
\label{SLP_PowerMin}
\mathbf{x}(\mathbf{s},\boldsymbol\gamma, \mathbf{H})&=&\arg\underset{\mathbf{x}}{\min}\|\mathbf{x}\|^2\\\nonumber
&s.t& |\mathbf{h}_j\mathbf{x}|^2\geq \gamma_j\sigma^2, \forall j\in K\\
&\quad&\angle(\mathbf{h}_j\mathbf{x})=\angle s_j, \forall j\in K.
\end{eqnarray}
The optimization can be tailored to exploit the detection region for any square multi-level modulation (i.e. $\mathit{M}$-QAM), the optimization can be formulated as:
\begin{eqnarray}\nonumber
\label{SLP_PowerMin2}
\mathbf{x}(\mathbf{s},\boldsymbol\gamma, \mathbf{H})&=&\arg\underset{\mathbf{x}}{\min}\|\mathbf{x}\|^2\\\nonumber
&s.t& \mathcal{R}\{\mathbf{h}_j\mathbf{x}\}\unlhd\sigma\sqrt{\gamma_j}\mathcal{R}\{s_j\},\forall j\in K\\
&\quad&\mathcal{I}\{\mathbf{h}_j\mathbf{x}\}\unlhd\sigma\sqrt{\gamma_j}\mathcal{I}\{s_j\}, \forall j\in K
\end{eqnarray}
where $\mathbf{x}\in\mathbb{C}^{N\times 1}$ is the output vector that modulates the antennas and $\unlhd$ is the operator that guarantees the signal is received at the correct detection region. This problem can be solved efficiently using second order cone programming \cite{Boyd}. It can be connected to  broadcast scenario (i.e. physical-layer multicasting \cite{Sidiropoulos:2006}), this connection has been thoroughly established and discussed in \cite{maha_TSP,Alodeh_TWC2}. 

Different symbol-level precoding schemes have been proposed in the literature. In \cite{maha_SPAWC,maha_twc}, the constructive interference precoding design is generalized under the assumption that the received MPSK symbol can reside in a relaxed region in order to be correctly detected. Moreover, a weighted maximization of the minimum SNR among all users is studied taking into account the relaxed detection region. Symbol error rate analysis (SER) for the proposed precoding is discussed to characterize the tradeoff between transmit power reduction and SER increase due to the relaxation. These precoding scheme achieve better energy efficiency in comparison to the technique in \cite{maha_ISIT}-\cite{maha_TSP}. In \cite{SLP_Korean_2014}, a symbol-level precoding scheme aims at manipulating both a desired signal and interfering signals is proposed such that the desired signal can be superimposed with the interfering signals. In this approach, a Jacobian-based algorithm is applied to improve the performance. Furthermore, it has been shown that robustness becomes stronger with an  number of co-scheduled users in the systems adopt $\mathit{MPSK}$ modulation.

Since the CSI acquisition in most systems is not perfect, it is important to design symbol-level schemes robust to different types of error. In \cite{SLP_Korean_2016}, interference is decomposed into predictable interference, manipulated constructively by a BS, and unpredictable interference, caused by the quantization error. To characterize performance loss by unpredictable interference, the upper bound of the unpredictable interference is derived. To exploit the interference, the BS aligns the predictable interference so that its power is much greater than the derived upper bound. During this process, to intensify the received signal power, the BS simultaneously aligns the predictable interference so that it is constructively superimposed with the desired signal. Different approach of guaranteeing the robustness of the symbol-level precoding is proposed in \cite{Masouros:2015,SLP_Korean_2014,Razavi_2015,SLP_Korean_2016}. These approaches are based on assuming that the errors in CSI is bounded, and the precoding  is designed taking into consideration the worst case scenario. The problem in \cite{Masouros:2015} is formulated as second order cone problem and can be solved using conventional convex optimization tools.

Most of the symbol-level precoding literature tackles the symbol-level precoding in single-level modulations (MPSK) \cite{Masouros2009,Masouros2011,Masouros:2015,SLP_Korean_2014,Masouros_THP,Masouros_khan,Masouros_2012,Masouros_security,Razavi_2015}. In \cite{maha_globecom,Alodeh_TWC2,Spano2016TSP}, the proposed precoding schemes are generalized to any generic modulation. The relation to physical-layer multicasting is established for any modulation in \cite{Alodeh_TWC2}. A per-antenna consideration is thoroughly discussed in \cite{Alodeh2016DSP}-\cite{Spano2016TSP}. In \cite{Spano2016TSP}, novel strategies based on the minimization of the power peaks
amongst the transmitting antennas and the reduction of the
instantaneous power imbalances across the different transmitted
streams is investigated. These objectives are important due to the per-antenna amplifiers characteristics which results in different amplitude cutoff and phase distortion. As a result, ignoring the previous factors can question the feasibility of employing precoding to multiuser MIMO systems. The work in \cite{Spano2016TSP} proposes to design the antenna weights taking into the account the amplifier characteristics by limiting the amount of power variation across the antennas amplifier, which leads to less deviation across the antennas and hence, less distortion.

The applications of symbol-level precoding span different research areas in wireless communications: underlay cognitive radio system \cite{maha_crowncom,Alodeh2016DSP,Masouros_2012,Masouros_khan}, coordinated multicell MIMO systems \cite{Ntougias_2016}, physical-layer security \cite{Kalantari:DM:ICASSP:2016,Kalantari:JSTSP,Masouros_security,SLP_Spoofing} and simultaneous wireless information and power transfer(SWIPT) \cite{Krikidis_SlP}. For more details about the applications of symbol-level precoding, please look at Table \ref{comparison1}.

Finally, symbol-leevel precoding and directional modulation is conceptually the same with the following main differences: directional modulation is driven by implementational aspects, assuming an analogue architecture with less emphasis on formulating criteria that optimizes the actual precoding weights. It also has less emphasis on  multiuser and system performance.
On the other hand, symbol-level precoding is driven by multiuser performance optimization, taking less consideration into implementation. However, it implicitly assumes a fully digital baseband implementation.

\begin{table}
	\begin{center}
		\centering
		\footnotesize\setlength{\tabcolsep}{2.5pt}
		\begin{tabular}{|p{1.5cm}|p{5.5cm}|}
			\hline
			Precoding& References\\
			\hline
			Block-level& Interference mitigation \cite{Per_Zetterberg,Karipidis2005CAMSAP,Karipidis2007,Karipidis2008,mats,Bengtsson1999,Peel2005,Wiesel_TSP_2008,Sadek_leakage}, \cite{Silva_TVT}  Energy efficiency \cite{mats,Bengtsson1999,Yu2007,Tervo_2015},\cite{Sidiropoulos:2006}, Fairness \cite{boche_mm,Karipidis2005CAMSAP,Karipidis2008,Christopoulos2014_TSP,Joudeh_Clerckx_SPAWC}, \cite{Christopoulos2014_TSP} \cite{Christopoulos2015_SPAWC}, Sum rate \cite{ghaffar} \cite{Christopoulos2014TWCOM} \cite{Finite_Alphabets_TWC}, Robust \cite{Jindal_imperfectCSI_MIMO,Shenouda2007,Robust_multigroup_multicasting_Icassp,Shen_2016}, Capacity\cite{multicast-jindal,caire_shitz_TIT}, Constant envelope\cite{Mohammed2013,Mohammed2013b} , Physical-layer security \cite{Kalantari_networking_coding,Security_BLP_chinese_Jsac,Liao_2011,MU_MIMO_security,Secure_SWIPT_2015}, SWIPT \cite{Secure_SWIPT_2015,Xu_2014_SWIPT,Multicasting_SWIPT}\\
			\hline
			Symbol-level& Energy efficiency \cite{maha_ISIT,maha_TSP,maha_SPAWC,maha_crowncom,maha_globecom,maha_twc,Alodeh_TWC2} Fairness \cite{maha_TSP},\cite{maha_twc}, Sum rate \cite{maha_TSP}, Robust \cite{Masouros:2015,SLP_Korean_2014,SLP_Korean_2016}, Interference exploitation \cite{Masouros2009,Masouros2011,Masouros_2012,Masouros_THP,Masouros_khan,Masouros_2012,maha_ISIT,maha_crowncom,maha_TSP,maha_SPAWC,maha_globecom,maha_twc,Spano2016Globecom,Alodeh_TWC2,Alodeh2016DSP,Spano2016TSP,SLP_Korean_2014,Masouros:2015,SLP_Korean_2016,Ntougias_2016,Amadori_2017}, Non-linear channels \cite{Spano2016Globecom,Spano2016TSP}, SINR balancing \cite{Masouros:2015}, Constant envelope \cite{Amadori_2017}, Physical-layer security \cite{Kalantari:DM:ICASSP:2016,Kalantari:JSTSP,SLP_Spoofing,Masouros_security},  Simultaneous wireless information
and power transfer (SWIPT)  \cite{Krikidis_SlP}\\
			\hline
		\end{tabular}
		\vspace{0.3cm}
		\caption{\label{comparison1} Precoding Classification Based on Switching Rate}
	\end{center}
\end{table}
\begin{table*}
	\begin{tabular}{|p{2.0cm}|p{2.50cm}|p{2.5cm}|p{9cm}|}
			\hline
			Precoding& Number of Groups& Number of Users&References\\
			\hline
			Broadcast& 1&$K$&Energy efficiency \cite{Sidiropoulos:2006}, Fairness (Capacity)\cite{Sidiropoulos:2006,multicast-jindal}, Robust\cite{Huang_2012} 	Physical-layer security \cite{Jeong_2014}, SWIPT\cite{Multicasting_SWIPT},Simplified \cite{Choi_2015}, Stochastic \cite{Rank_two_Ma} \\
			\hline
			Unicast&$K$&1& Interference mitigation\cite{Per_Zetterberg,Peel2005,Sadek_leakage,Lai-U:2004,Spano2016TSP,Wiesel_TSP_2008,Kai_Kit_Wong_2003,Shen_2016}, Interference exploitation \cite{Masouros2009,Masouros2011,Masouros_khan,Masouros_2012,Masouros_THP,Alodeh2016DSP,Alodeh_TWC2,Spano2016TSP}, Energy efficiency\cite{mats,Bengtsson1999,boche,Yu2007,Tervo_2015,Masouros_perturbation_2014}, Fairness \cite{boche_mm}, Finite alphabets \cite{Finite_Alphabets_TWC,Zhi_Ding_TWC,ghaffar}, Robust \cite{Shenouda2007,Robust_feifei_second_order,Boche_Robust,Robust_MMSE,Wang_Palomar_worst_case,Huang_2012,Ma_2017,Chinese_robust_probablistic,Joudeh_Clerckx_TSP,Sohrabi_2016,Medra_2016}, Robust interference exploitation\cite{SLP_Korean_2016,SLP_Korean_2014,Masouros:2015}Constant envelope\cite{Mohammed2013,Mohammed2013b,Amadori_2017}, Per-antenna optimization \cite{Yu2007,Shen_2016,Spano2016TSP}\\
			\hline
			Multicast&$G$&$K/M$&Fairness\cite{Karipidis2005CAMSAP,Karipidis2007,Karipidis2008,Christopoulos2014_TSP,Christopoulos2014TWCOM,Christopoulos2015_SPAWC}, Energy efficiency \cite{oskar_ICC}, Interference mitigation \cite{Silva_TVT}, Robust  \cite{Robust_multigroup_multicasting_Icassp,Robust_two_rank_multigroup_multicasting}, Stochastic \cite{Wu_amplify_2016}, Coordinated \cite{Xiang_coordinated}, Relay \cite{Bornhorst_2012}\\
			\hline
		\end{tabular}
		\vspace{0.2cm}
		\captionof{table}{\label{comparison2} Precoding Classification Based on the Group Size }
\end{table*}
\section{Comparative Study}\label{Comparative_Study}
In order to assess the relative performance of the precoding techniques discussed in the previous sections, some numerical results are presented in this section. Firstly, the focus is on block-level precoding, both unicast and multicast. Then, the performance of symbol-level precoding is assessed, in comparison to the conventional block-level case.

In the remainder of this section, a system with 4 transmit antennas and 4 users is assumed, hence having  $N = K = 4$. Moreover, the channel vector of the generic user $j$ is modeled as $\mathbf{h}_j \sim \mathcal{CN}(0,\sigma_h^2\mathbf{I})$, with $\sigma_h^2 = 1$ and the results are obtained averaging over several channel realizations. Furthermore, we assume a unit AWGN variance for all the users. 

\subsection{Block-level Precoding Results}
\begin{figure}[tbp]
\centering
\includegraphics[width=1\columnwidth]{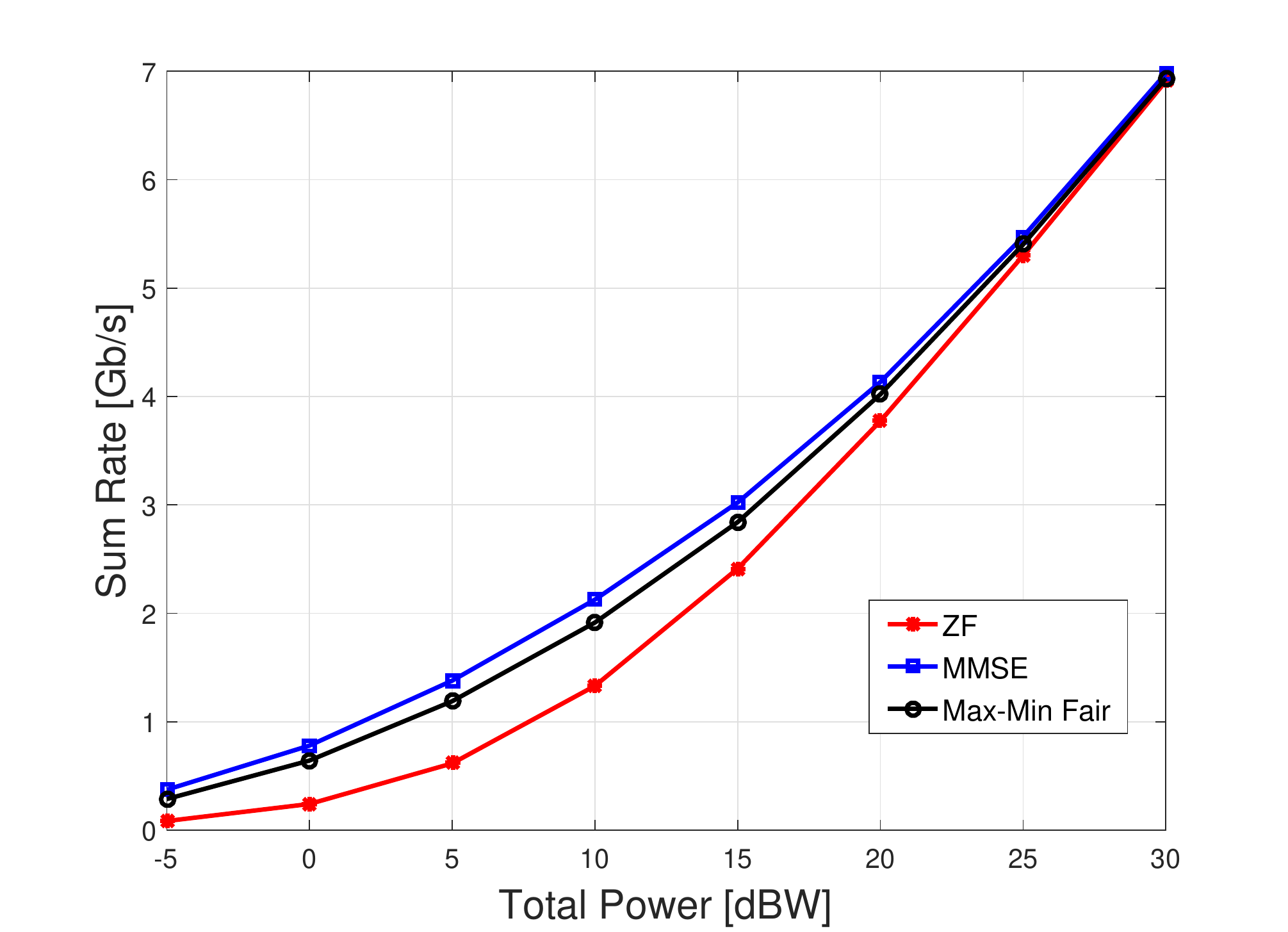}\\
\caption{Sum rate of different unicast block-level precoding, in Gb/s, versus total available power, in dBW.}\label{fig:rate_unicast}
\end{figure}

Considering a unicast framework, Fig. \ref{fig:rate_unicast} compares the sum-rate performance of ZF precoding, MMSE precoding, and the max-min fair scheme given in \eqref{max_min_SPC_Unicast}. A system bandwidth of 250 MHz is assumed for the rate calculation. Interestingly, the best performance is given by MMSE.

\begin{figure}[tbp]
\centering
\includegraphics[width=1\columnwidth]{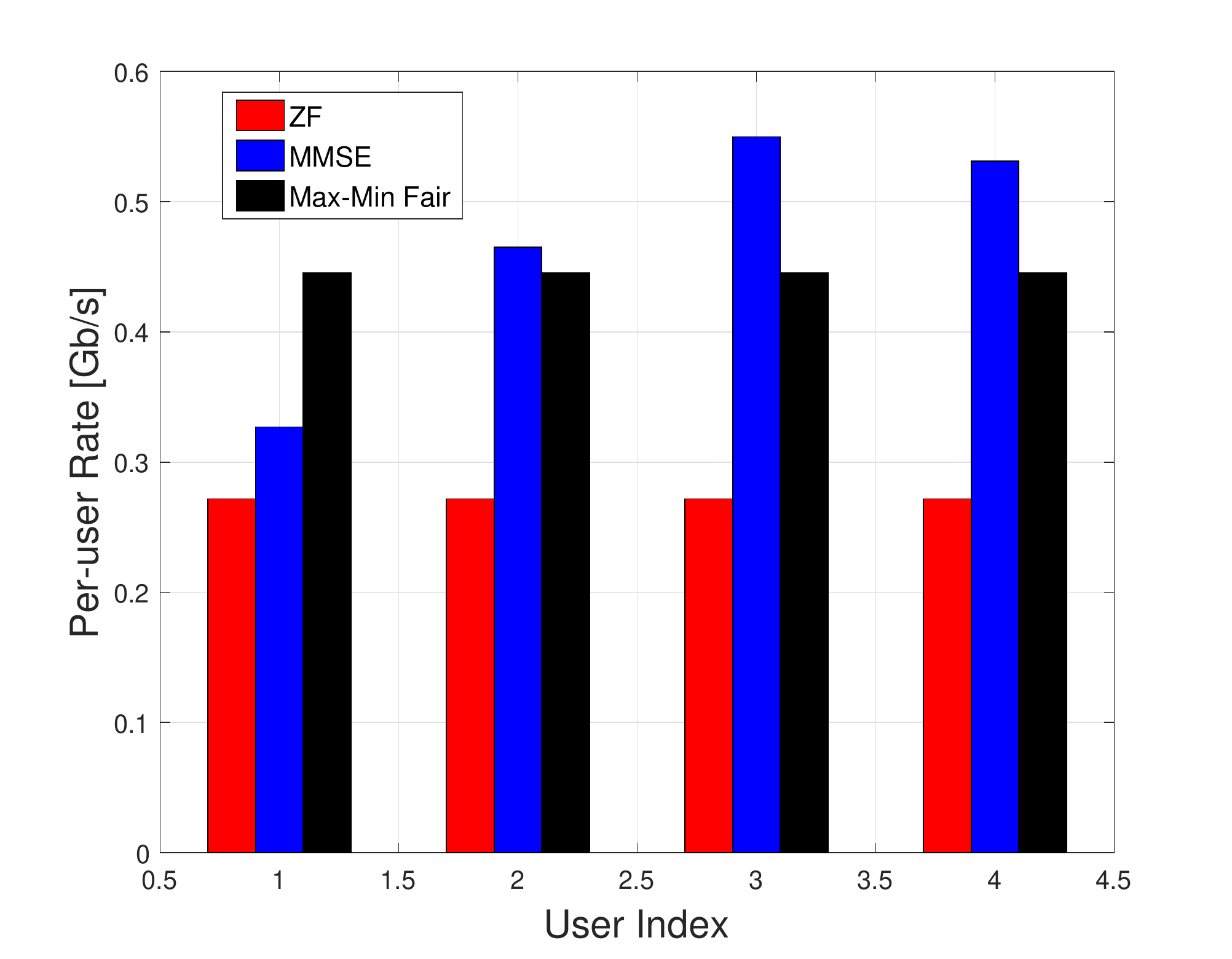}\\
\caption{Per-user rate distribution, in Gb/s, versus total available power, for a specific channel realization.}\label{fig:bars_unicast}
\end{figure}

Furthermore, Fig. \ref{fig:bars_unicast} shows how the sum rate is distributed among the users for a specific channel realization.
Although the max-min fair approach performs slightly worse than MMSE in terms of sum rate, it is visible how it guarantees a better minimum rate across the users. Therefore, it improves the fairness.

\begin{figure}[tbp]
\centering
\includegraphics[width=1\columnwidth]{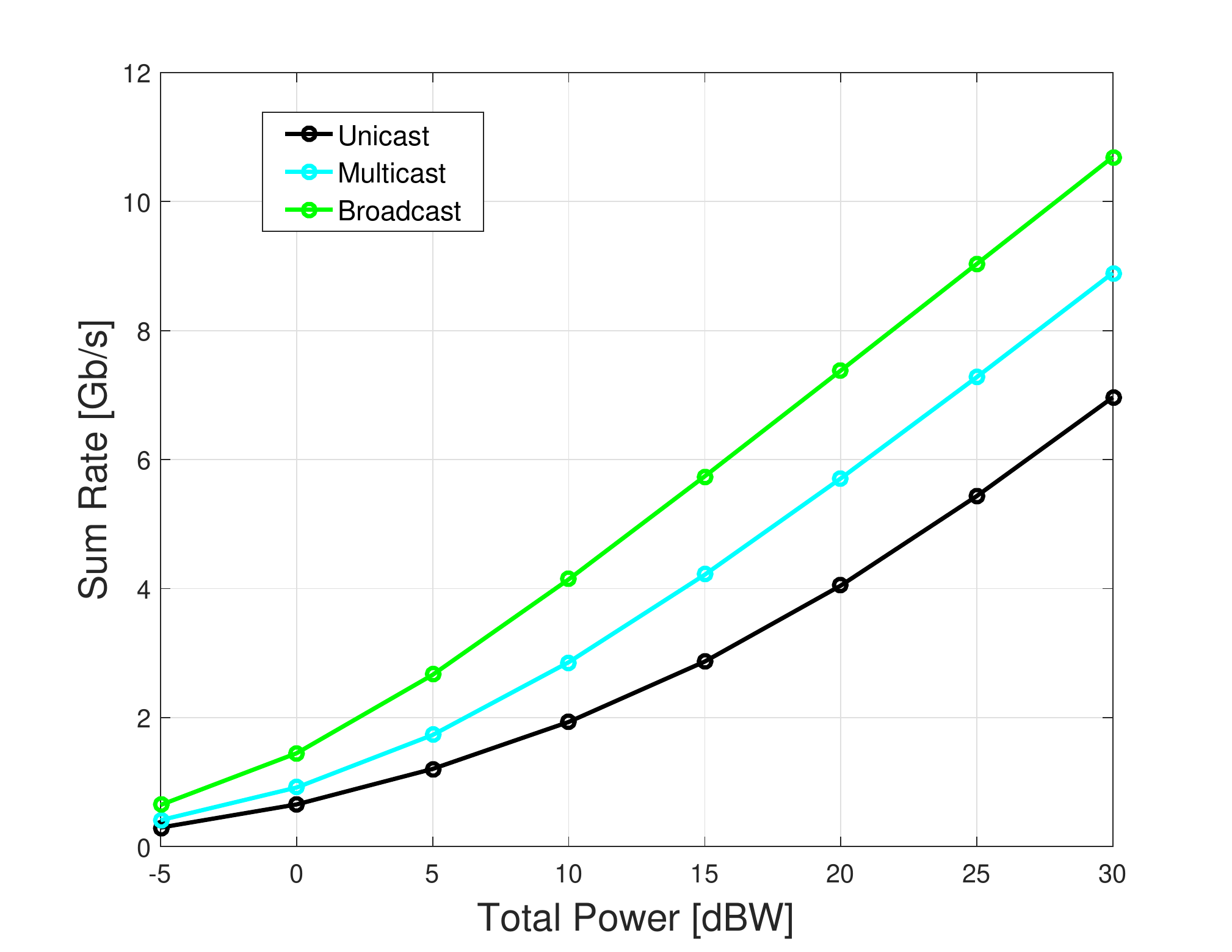}\\
\caption{Sum rate performance of block-level max-min fair for different service types, in Gb/s, versus total available power, in dBW.}\label{fig:rate_multicast}
\end{figure}

\begin{figure}[tbp]
\centering
\includegraphics[width=1\columnwidth]{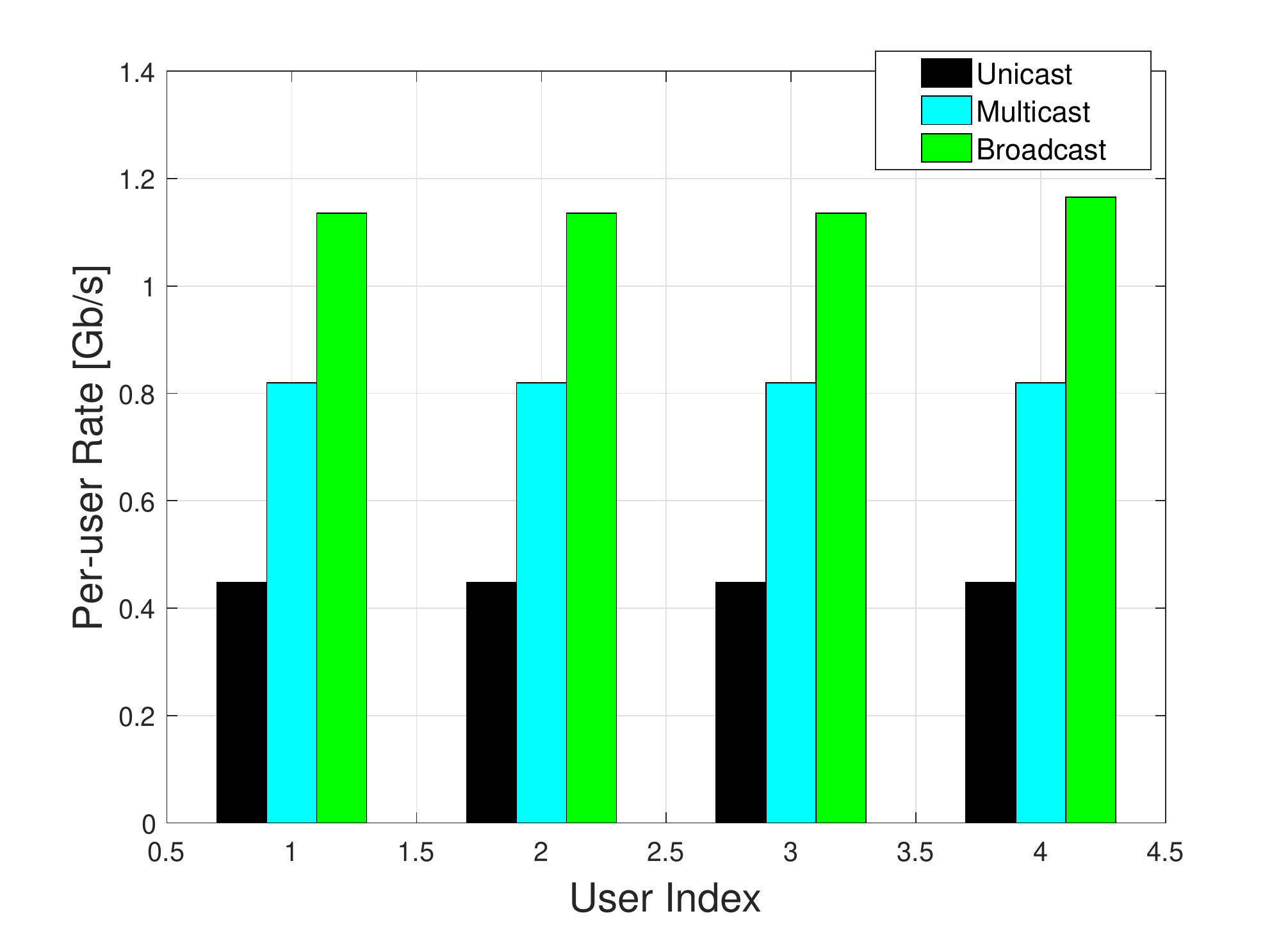}\\
\caption{Per-user rate distribution, in Gb/s, versus total available power, for a specific channel realization.}\label{fig:bars_multicast}
\end{figure}

We consider analogous numerical results for comparing the introduced precoding techniques for unicast, multicast, and broadcast (the max-min fair optimization strategy is considered). Fig. \ref{fig:rate_multicast} displays the sum rate as a function of the total available power. It emerges how the performance improves when different users are grouped so as to receive the same data stream. This can be justified by the fact that in the multicast case the interference is reduced with respect to the unicast case, where each user receives a different stream. The same can be noticed from the result of Fig. \ref{fig:bars_multicast}, where the rate distribution is shown for the three cases considering a specific channel realization.

\subsection{Symbol-level Precoding Results}
In this section, we compare the performance of symbol-level precoding with the equivalent block-level precoding scheme, in a unicast framework. In particular, we consider the power minimization strategy with $\mathrm{QoS}$ constraints, given in \eqref{SPM_Unicast} and in \eqref{SLP_PowerMin} for block-level and symbol-level respectively. A 8-PSK modulation scheme is assumed for the data information. 

\begin{figure}[tbp]
\centering
\includegraphics[width=1\columnwidth]{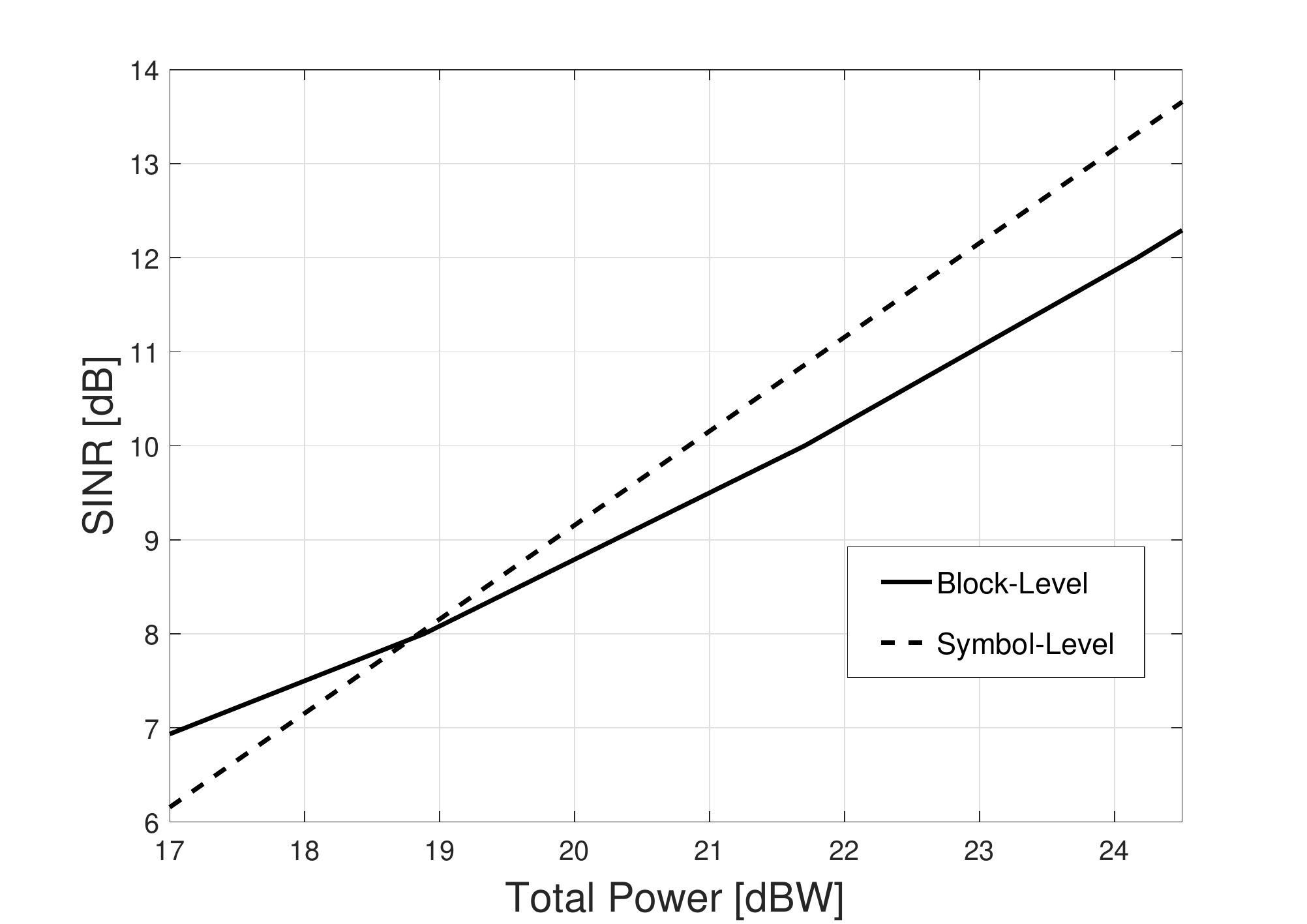}\\
\caption{Average attained SINR of block and symbol-level, in dB, versus total available Power, in dBW, for a 8-PSK modulation Scheme.}\label{fig:SINR_slp_block}
\end{figure}

Fig. \ref{fig:SINR_slp_block} shows the related performance obtained for the two schemes, in terms of attained average SINR, as a function of the required total power. It is clear how the symbol-level precoding scheme outperforms the block-level one in the high SINR regime. This can be justified by considering that this regime, which corresponds to a higher transmitted power, is more interference limited. Accordingly, the symbol-level scheme can leverage the interference to improve the overall performance. 

\begin{figure}[tbp]
\centering
\includegraphics[width=1\columnwidth]{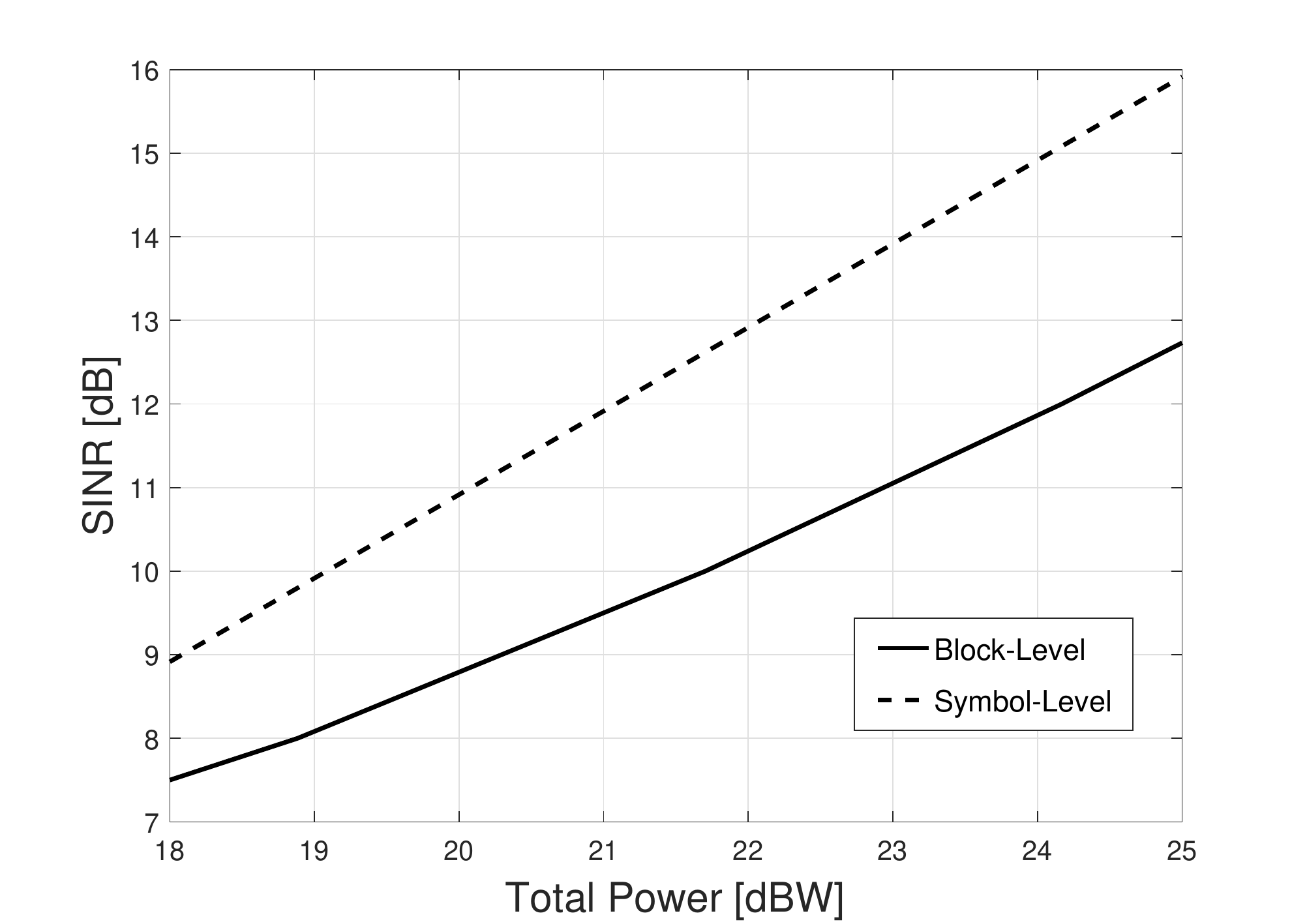}\\
\caption{Average attained SINR of block and symbol-level, in dB, versus total available power, in dBW, for a 16-QAM modulation Scheme.}\label{fig:SINR_slp_block_16QAM}
\end{figure}

Fig. \ref{fig:SINR_slp_block_16QAM} shows an analogous comparative result considering a multi-level modulation. In particular, a 16-$\mathit{QAM}$ modulation scheme is used for the data information. It is clear how symbol-level outperforms the block-level precoding for all available power values. From Fig.\ref{fig:SINR_slp_block}-\ref{fig:SINR_slp_block_16QAM}, it can be concluded that the modulation type plays an important role in symbol-level precoding systems. The rectangular modulations (MQAM) outperform the circular modulation (MPSK and APSK) due to the relaxed detection region of rectangular modulations.

\section{Challenges and Open Problems}\label{Open Problems}
\subsection{Robust Precoding}
The accuracy of the estimated CSI plays
an important role in designing accurate precoding that can
mitigate and exploit the interference created from the simultaneous spatial transmissions. Designing robust precoding strategies is an important topic to tackle, especially when the acquired CSI is not perfect \cite{Shenouda2007,Boche_Robust,Robust_MMSE,Robust_feifei_second_order,Robust_two_rank_multigroup_multicasting}.For these cases, robust precoding  under uncertainty is required. In this direction, three robust different designs were proposed in the literature\cite{Gershman_Convex}. Namely, the probabilistic design \cite{Chinese_robust_probablistic}, where acceptable performance is guaranteed for some percentage of time, the expectation based design that requires knowledge of the second order channel statistics but cannot guarantee any outage performance \cite{Robust_feifei_second_order} and the worst-case design\cite{Wang_Palomar_worst_case}. The latter approach guarantees a minimum $\mathrm{QoS}$ requirement  for any error realization. 

Most of the techniques in the literature focus on designing robust strategies for block-level precoding. For symbol-level techniques, there is room to design robust strategies to tackle different types of uncertainties, since the only proposed robust design tackles worst case for single-level modulation \cite{Masouros:2015}. Robust strategies tackling different kinds of uncertainties for multi-level modulation still need to be addressed to see the full potential of symbol-level precoding.   \\

\subsection{Multiple-antenna Users 	Terminals}
Having multiple antenna users' terminal adds a new dimension that can be utilized in different ways. 
Most of the literature focuses on exploiting them in unicast block level precoding \cite{Jindal_combining_2008,Jindal_imperfectCSI_MIMO,Trivellato_JSAC_MIMO,Heath_Andrews_TSP_MIMO,Joham,Khandani_TIT,Lizhong_Zheng_2003,Lozano_2010,emil_MIMO_TSP_RC}. Three methods have been proposed in the literature to use these additional DoF: receive combining \cite{Jindal_combining_2008}, multistream multiplexing, and receive antenna selection. All these schemes have their advantages in comparison to single-antenna receiver. In \cite{Jindal_combining_2008}, receive antenna combining has been used to reduce channel quantization error in limited feedback MIMO downlink channels,
and thus significantly reducing channel feedback requirements. In \cite{Trivellato_JSAC_MIMO,Heath_Andrews_TSP_MIMO,Joham,Khandani_TIT,Lizhong_Zheng_2003,Lozano_2010,emil_MIMO_TSP_RC}, different contradicting conclusions are drawn related to multistream spatial multiplexing.
The authors of \cite{Trivellato_JSAC_MIMO} claim that transmitting at most one
stream per user is desirable when there are many users in
the system. They justify this statement by using asymptotic
results from \cite{Khandani_TIT} where $K\rightarrow\infty$. This argumentation ignores
some important issues: 1) asymptotic optimality can also be
proven with multiple streams per user; 2) the analysis implies
an unbounded asymptotic multi-user diversity gain, which is
a modeling artifact of fading channels \cite{Dohler_com_mag}. The diversity-multiplexing tradeff (DMT) brings
insight on how many streams should be transmitted in the
high-SNR regime \cite{Lizhong_Zheng_2003}, \cite{emil_MIMO_TSP_RC} considers how a fixed number of
streams should be divided among the users.

In the context of this survey, the utilization of multiple antenna at receiver can achieve potential gains and open the doors to new problems that can be solved as discussed below. 
\subsubsection{Symbol-level Precoding}
The exploitation of multi-antenna at the receivers has never been discussed for symbol-level precoding. It is interesting to explore the potential gains that can be achieved if we use the different schemes. In \cite{Alodeh_TWC2}, simulations have shown that required power to achieve the requested $\mathrm{QoS}$ decreases with system size. It is  interesting to see how the system will behave if we have multi-stream spatial multiplexing, what is the optimal number of streams per user? Is it modulation dependent? Can the performance show some gains if we have diversity in the system? DMT analysis is required to investigate the system performance at high SNR regime. Moreover, it is worth to see if the different receive antennas selection or receive combining algorithms proposed for the unicast block-level precoding are applicable for symbol-level precoding, do we need new algorithms to achieve unprecedented gains in  symbol-level precoding.
\subsubsection{Multicast Precoding}
Adding multiple antenna at the receiver can be beneficial for multicast precoding. There is no deep investigation to optimize the multiple antenna at the receiver side. There are still open problems that need to be addressed. Different questions need to be solved: Can developing a new receive combining to optimize the performance be different from block-level unicast approaches? Can the user's group affect the optimal receive combining strategy? Can any user belong to more than one group to receive multiple streams simultaneously using multistream spatial multiplexing? In the literature, the optimal group size has been investigated in \cite{Matskani}. DMT analysis is required to study the multigroup multiplexing gains and diversity, and what is the optimal number of the groups and the optimal size of each group with respect to the number of BS 's antennas. 

\subsection{Multicast Symbol-level Precoding}
Symbol-level precoding has been applied so far in unicast precoding to exploit the intererence among the different data streams. However, it has the potential to treat inter-group interference in multicast scenarios. The challenge in this direction is to properly exploit the constructive interference when the number of targeted users is larger that the number of transmit antennas. The importance of solving this problem lies on the framing structure of communications standards, where
each frame is received and decoded by all co-group. Therefore,
the same precoder should be applied to all users served by the same frame.  

 Since
each frame is received and decoded by all co-group users,
the design of an optimal frame based precoder is given by
solving a multicast multigroup optimization problem. Thus,
multicasting allows for an analytically formal modeling of the
problem. Therefore, in the context of frame based precoding,
the fact that the same precoder needs to apply to the different
data of many receivers due to the framing constraint, leads to
a multicast consideration.
\subsection{Symbol-level Precoding Side-Effects on other Blocks of the Communication Chain}
\subsubsection{Precoded Pilots for SNR Estimation in Symbol-level Precoding}
These functions are often neglected in precoding studies, but they are crucial in implementing a novel precoding method. 

Focusing on SNR estimation, this presents a challenge when symbol-level precoding is used. The reason is that unless per-user SNR constraints are imposed the instantaneous received power at each user ranges depending on the input symbol vector. In general, the block-level SNR can be estimated by averaging over a large number precoded input symbol vectors. However, due to the pilot overhead the number of input symbol vectors that can be utilized is limited. The challenge here is to devise pilot design techniques that can reliably estimate the average SNR with a limited pilot length.  

\subsubsection{Modulation and Coding Allocation}
Focusing on modulation and code allocation and scheduling, these are important functions which raise cross-layer issues between the physical, MAC and network layers. More specifically, the modulation and coding allocation delimits the achieved rates at each user. The modulation and coding is assigned per user based on the predicted average SNR over a symbol block.  In conventional block-level precoding, this average SNR can be efficiently calculated at the transmitter given the scheduled set of users. However, in symbol-level precoding the calculation is more complex since it has to be calculated symbol-by-symbol and averaged over a statistically important set of symbol vectors. Computational-efficient heuristic methods for this process is an important open topic. It should be noted that a workaround is for the users to feedback the requested rates to the transmitter as suggested in \cite{Alodeh_TWC2}.

\subsection{Massive MIMO}
Massive MIMO (also known as “Large-Scale Antenna Systems”) is an emerging technology, that scales up MIMO by possibly orders of magnitude to utilize the huge spatial multiplexing gains \cite{Bjornson_Massive,Swindlehurst_one_bit_analysis,Rusek_Massive,Massive_Larsson_one_bit,Massive_MIMO_waveform_ahmet,massive_tao,Bjornson_massive_non_ideal}. The basic premise behind massive MIMO is to glean all the benefits of conventional MIMO, but on a much
greater scale. The anticipated huge spatial multiplexing gains (degrees-of-freedom DoF) are achieved by coherent processing over large antenna arrays, which result in strong signal gains, low interference, reduced latency, and robustness to imperfect channel knowledge. This comes at the expense of infrastructure costs; the hardware requirements and circuit power consumption scale linearly with the number of BS antennas $N$. In contrast to the current systems, with conventional expensive and power-hungry BS antenna circuits, the main key to cost-efficient deployment of large arrays is low-cost antenna circuits with low power consumption.  The challenge is to make many low-precision components work effectively together. Such low-cost transceivers are prone to hardware imperfections, but it has been conjectured that the huge degrees-of-freedom would bring robustness to such imperfections. 
Another challenge in massive MIMO systems is the CSI acquisition \cite{Contamination,Contamination_jose,Contamination_ALODEH,Contamination_Yin,Contamination_Rami,Contamination_Surveys}. In the literature, different acquisition techniques in time division duplexing (TDD) and frequency division duplexing (FDD) are proposed.
The applications of symbol-level and multicast precoding to massive MIMO are discussed below:
  
\subsubsection{Symbol-level Precoding}
In the context of this paper, symbol-level precoding can be a good candidate to be utilized in massive MIMO system. The premise of having a transmitter equipped with many more antennas than the number of served users can produce an excess of degrees of freedom. These additional DoFs could be potentially exploited in symbol-level precoding  to improve the conventional  
performance metrics, but also to further shape desirable waveform properties such as peak to average power ratio (PAPR)
and spectral characteristics. This opens the doors to a very promising direction, since it might entail more cost-efficient and less complex transmitter architectures. Moreover, the impact of having limited CSI on the waveform design for massive MIMO is still an open problem that needs to be addressed.
\subsubsection{Multicast Precoding}
In multicast precoding, a common assumption is that the number of served users is higher than the number of transmit antennas. However, the premise of massive MIMO can overcome this assumption and enable a different view of multicast precoding. When the number of transmit antennas is abundant, we might be able to enable multi-rank transmission to each group to serve efficiently more  users, especially when they belong to the same group but they have semi-orthogonal channels.   
\subsection{Millimeter Wave}
The spectrum congestion in frequencies that are already occupied for mobile services along with the
enormously increasing demands for mobile services, forces the wireless communications industry
to explore systems adapted to frequencies within the so-called millimeter Wave (mmWave) band
\cite{Rappaport2013,Rangan_2014}. The development of such mmWave transceivers is a very
challenging task. Due to their nature, mmWave signals suffer from severe degradations though,
due to the short wavelength of mmWave frequencies, a prospect transceiver may employ large array structures for providing high beamforming gains or improvements in the system’s spectral efficiency via precoding techniques. Existing digital pre/post-coding techniques, developed in the past years for lower frequency MIMO systems are not suitable for
systems of large antenna arrays due to high demands in hardware complexity and power consumption. This is the case since a fully digital transceiver requires a dedicated Radio Frequency (RF)
chain per antenna which includes a number of different electronic elements (e.g., Digital-to-fAnalog/Analog-to-Digital converters) that are of high hardware complexity and power
consumption, especially for large antenna arrays. 

Thus, recent literature approaches seek for solutions that are based on transceivers employing only few number of RF chains compared to the number of antennas and apply hybrid analog-digital precoding to optimize the transmission 
\cite{XZhang2005,Ayach2014}. The latter techniques are based on a two stage precoder, a digital one applied in the baseband domain and an analog one, applied in the RF domain via a network of phase shifters. While a number of different works 
\cite{Dai2016,Alkhateeb2016,Tsinos2016a,Tsinos2016b} were developed in the past in the context of hybrid precoding with satisfactory performance, it is possible that in several cases their implementation may still be of high complexity and power consumption \cite{Tsinos2016c}. From that point of view, it is highly desirable to reduce the complexity and power consumption as much as possible, that is used by transceivers of a single RF chain, e.g. by removing completely the digital counterpart of the hybrid precoder. Unfortunately, such a RF-only beamformer can support only single stream and very primitive multi-user communications resulting in severe performance losses. This is the point were directional modulation aims at stepping in to provide efficient precoding schemes for single RF chain transceivers to support multiple streams. 

From the discussion given in Section IV.A, directional modulation techniques develop symbol level precoding directly in the RF domain via digitally controlled analog components (e.g. phase shifters and attenuators). However, there are several challenges toward the implementation of a fully functional and efficient transceiver related to the impairments on the analog hardware that could result in severe performance degradation, efficient CSI estimation techniques, since there is no straightforward way to estimate the required information and waveform design aspects. 

Furthermore, as it was discussed in Section III.A, broadcast precoding techniques were recently examined for hybrid analog digital transceivers \cite{Choi_2015,Dai2016,Demir2016}. Due to the increasing interest around hybrid or in general solutions that exhibit low complexity in large array systems, such as the mmWave or massive MIMO ones, several digital communications techniques have to re-examined in order to propose solutions that can be applied efficiently in the latter systems. Thus, the numerous digital techniques developed in the context of broadcast precoding during the past years, could be examined towards that direction providing new challenges and rekindle the interest in this well-studied field.  
  

\section{Conclusions}\label{conclusion}

The integration of multiuser MIMO/MISO has been considered in different standards such as LTE/LTE-Advance \cite{Li_2014,clerckx_wimax,Lee_2012_LTE,Lim_2013}, and IEEE 802.11ac \cite{Bejarano_2013}. In the era of heterogeneous networks, there are many challenges to overcome in the context of multiuser MIMO to achieve better resource utilization.

In this context, this survey classified the multiuser MIMO precoding strategies 
 with respect to two major axes: the number of users addressed by each information stream, and the switching rate of the precoding coefficients. 
According to the first classification criterion, unicast, multicast, and broadcast precoding strategies have been throughly discussed. To achieve the optimal network efficiency (throughput, energy efficiency, delay, ...etc), an optimized combination of these transmission strategies can be the new interface for the next wireless generation.

With respect to the second classification criterion, i.e., the switching rate, block-level precoding and symbol-level precoding schemes have been considered. While the former class refers to the conventional schemes, whereas precoding exploits the CSI and is applied over symbol blocks, the latter class refers to novel techniques applying precoding on a symbol-by-symbol basis, thus able to exploit the data information, together with the CSI, in the signal design.  We introduced directional modulation and symbol-level precoding for constructive interference where they share the same conceptual model, designing the antenna weights on symbol by symbol basis. However, the directional modulation focuses on the implementational aspects of the concept while the symbol-level precoding focuses on the multiuser optimization aspect. Some representative optimization strategies for symbol-level precoding have been discussed, both for single-level and multi-level modulation schemes. Despite the fact that symbol-level precoding techniques seem to be futuristic since they incur huge computational complexity at the base station, it can be argued that computational complexity can be transferred to the cloud RAN level \cite{Checko_2015}. 

In order to assess the performance of the presented precoding schemes, some numerical results have been presented in a comparative fashion, in terms of attained rate and SINR at the receivers' side. In the context of block-level precoding, the results highlight how the optimization-based schemes outperform the closed-form solutions with respect to specific targeted objectives, e.g., the fairness amongst the users. Moreover, it emerged how, by applying the proper precoding schemes, the multicast framework can show better performance than the unicast one, given a fixed total number of users. Furthermore, it has been shown how symbol-level precoding outperforms the corresponding block-level scheme in interference limited regimes.

Finally, a number of open challenges related to the presented precoding techniques are discussed, so as to pave the way to the future research in this promising area.

\end{document}